\documentclass[10pt]{iopart}

\expandafter\let\csname equation*\endcsname\relax
\expandafter\let\csname endequation*\endcsname\relax
\usepackage{cite}
\usepackage[usenames,dvipsnames]{color}
\usepackage{bbm}
\usepackage[english]{babel}
\usepackage[utf8x]{inputenc}
\usepackage[T1]{fontenc}
\usepackage{amsmath,amssymb}
\usepackage{graphicx}
\usepackage{xcolor}
\usepackage{float}
\usepackage{stackrel}
\usepackage{amsfonts}
\usepackage{bm}
\usepackage[normalem]{ulem}
\usepackage{hyperref}
\usepackage{verbatim}
\usepackage{physics}
\usepackage{caption}

\newcommand{\beq}{\begin{equation}}
	\newcommand{\eeq}{\end{equation}}
\newcommand{\eq}{\begin{equation}}
	\newcommand{\en}{\end{equation}}

\begin{document}
	
	\title{Spatial correlations and entanglement in a hybrid system of $N$ fermion pairs with harmonic interaction}
	
	\author{M. D. Jim\'{e}nez$^{1}$, W. J. Díaz$^{2}$, E. Cuestas$^{3,4}$, A. Valdés-Hernández$^{5}$, and A. P. Majtey$^{1}$} 
	\address{$^{1}$ Instituto de F\'{i}sica Enrique Gaviola, CONICET and Universidad Nacional de C\'{o}rdoba, Ciudad Universitaria, X5016LAE, C\'{o}rdoba, Argentina.}
	\address{$^{2}$ Universidad Nacional de C\'{o}rdoba, Facultad de Matemática, Astronomía, Física y Computación, Ciudad Universitaria, X5016LAE, C\'{o}rdoba, Argentina.}
	\address{$^{3}$ OIST Graduate University, Onna, Okinawa, Japan}
	\address{$^{4}$ Forschungszentrum Jülich GmbH, Peter Grünberg Institute, Quantum Control (PGI-8), 54245 Jülich, Germany}
	\address{$^{5}$ Instituto de F\'{\i}sica, Universidad Nacional Aut\'{o}noma de M\'{e}xico, Apartado Postal 20-364, Ciudad de M\'{e}xico, Mexico.}	
	\ead{martin.jimenez.388@mi.unc.edu.ar}
	\date{\today}
	
	\begin{abstract}
		Using the Moshinsky model, we  analyze the spatial correlation and the entanglement of the ground state across different bipartitions of a system composed by $N$ pairs of harmonically confined fermions of two different interacting species. We find that in the strongly attractive regime fermions tend to localize within a confined region, while Pauli exclusion induces a spatial repulsion among identical particles. Conversely, in the strongly repulsive regime, the system exhibits phase separation into two spatially distinct domains. We propose a suitably designed entanglement measure that takes into account the (in)distinguishable nature of the particles, so as to  guarantee that only quantum correlations beyond exchange or Slater correlations contribute to the entanglement. Our findings reveal how entanglement varies across different bipartitions, influenced by both the number of pairs and the interaction strength. 
		The obtained entanglement between distinguishable species increases with both $N$ and the interaction strength, whereas the entanglement between subsystems of which at least one contains particles of both species decreases with $N$, suggesting a screening effect that weakens effective interactions among parties of the same kind. Our results sheds light on the intricate interplay between particle statistics and entanglement dynamics in composite quantum systems, and provides insights into the fundamental quantum correlations in fermionic systems.
	\end{abstract}
	
	\maketitle
	\pagestyle{plain}
	\section{Introduction}
	
	The understanding and description of the behavior of interacting quantum many-body systems, and the design and exploitation of these systems for quantum information tasks, are some of the most important challenges in quantum physics. Quantum systems consisting of many particles exhibit a wide variety of physical phenomena, some of which can lead to fascinating (and promisingly useful for technological applications) macroscopic properties such as high-temperature superconductivity or thermalization \cite{zhou_2021,mori_2018,abanin_2019,Annett_book}. The primary challenge when dealing with quantum many-body systems arises from the fact that in the majority of cases the Schrödinger equation is not amenable to analytical solution. Even when employing numerical techniques, the determination of the wave function remains a significant hurdle. Solving most of these systems is extremely difficult, and it is essential to develop techniques that improve our understanding of the exhibited phenomena. On the other hand, solvable models give us a way to understand quantum dynamics that would otherwise be inaccessible. They enlighten us with relevant information about the behavior of correlation functions, the evolution of entanglement, and other fundamental and practical aspects, see for example Refs. \cite{armstrong_2011,moshinsky_1968,moshinsky_1968_errata,ebrahimi_fard_2012}. 
	
	A system of particular interest, since it admits an analytical solution, is the so-called Moshinsky model, consisting of two particles in a harmonic trap coupled by a harmonic interaction \cite{moshinsky_1968,moshinsky_1968_errata,moshinsky_book_ho}. Because the model and its analytical solution are generalizable to practically any number of particles it has become a testing ground for the study of various systems. 
	For example, the $N$-harmonium system, which is a completely integrable model of $N$ particles where both the confinement and the two-particle interaction are harmonic, has provided further insight into numerous phenomena in a variety of physical systems ranging from nucleus to molecules \cite{johnson_1991,amovilli_2003,cuestas_2017,cuestas_2020}. It has also been implicitly used to study cold atoms \cite{Armstrong2011}, and more recently the thermodynamic properties of a pair-interacting heavy hole gas confined within a lens-shaped Ge/Si quantum dot, were investigated using the two-dimensional Moshinsky model to represent the pair interaction potential as an oscillator function dependent on the interparticle distance \cite{Mkrtchyan2024}. 
	Additionally, mutual information has been studied to analyze  correlations in $N$-particle systems, both in terms of the interaction strength and the particle number, within Moshinsky-like oscillator models \cite{peng2015statistical,salazar2024mutual}.
	
	When dealing with multipartite quantum systems, a most relevant  phenomenon that deserves attention is entanglement, a type of quantum correlation with no classical counterpart, that stands as a fundamental resource in many quantum information processing tasks \cite{acin_2018, awschalom_2022}. Further, when the constituents of the system are indistinguishable particles, exchange correlations emerge naturally. Whether these correlations contribute to the useful entanglement between the parties has been a subject of debate \cite{killoran_2014,becher_2020}. According to the approach followed here, a state of indistinguishable fermions is separable (or non-entangled) if and only if it has Slater rank 1, which in the case of a pure state means that the state vector can be obtained from the antisymmetrization of a product state. Based on this definition of entanglement in systems of identical fermions, simple necessary and sufficient separability criteria have been established \cite{schielmann_2001,pauskaukas_2001,plastino_2009_epl,Majtey2023,Benatti_2020}. In particular, using those definitions, the ground-state entanglement of the $N$-harmonium system was analytically determined in terms of $N$ and the relative interaction strength \cite{benavides2014entanglement}. 
	
	In the present contribution we focus on the study of the entanglement and spatial correlations in the many-body ground state of a system composed by $N$ identical fermions of species $a$ and $N$ identical fermions of species $b$ with harmonic interactions among fermions of different species, and, importantly, with any two fermions of different species being distinguishable from each other. 
	The distinction between two or more fermionic species arises in many physical scenarios, for instance, when each species occupies a well-defined spatial location, as occurs in discrete lattices, or when the particles possess distinct physical properties such as mass, atomic number, or spin \cite{pecak_2016,bouvrie_2019}. 
	The need for a hybrid model emerges naturally in a system where the $a$-type fermion represents a spin-up state and the $b$-type fermion corresponds to a spin-down state, particularly in the absence of spin-flip processes \cite{Cuestas2022,zurn_2012}. 
	An important physical realization of this scenario is found in dilute Fermi gases in the quantum degenerate regime \cite{giorgini_2008_review}. Due to the low temperature and large mean interparticle distance the dominant interactions in ultracold Fermi gases are governed by $l=0$ scattering processes. In turn, the antisymmetry requirement for the wave function of identical fermions excludes $s$-wave scattering between spin polarized particles, whereas $s$-wave interactions between fermions of opposite spin remain allowed and can even be tuned using Feshbach resonances. This motivates our choice of a model where only fermions of different species interact and also justifies the consideration of an unpolarized system, i.e. a system with the same amount of particles of different species. The BEC-BCS crossover in ultracold Femi gases also allows for another physical realization of this two-species particle system, namely, molecular condensates made up of atoms with different hyperfine states \cite{giorgini_2008_review, bouvrie_2019}. While our use of harmonic potentials is clearly idealized, it allows for an analytically tractable setting in which fundamental aspects of spatial correlations and entanglement can be studied in full detail.
	Moreover, the second order harmonic approximation for generic potentials in combination with a two-species model has been proven useful to address Wigner Molecules and to capture the Friedel-Wigner transition \cite{cuestas_2020}. We analytically compute the relevant reduced density matrices that allow us to address the spatial correlations of certain subsystems of interest as well as the entanglement across different bipartitions. In particular, we propose an entanglement measure that generalizes previous definitions \cite{majtey_2016,plastino_2009_epl} to this hybrid system by excluding the exchange correlations among indistinguishable parties, and that reduces to the standard measures of entanglement when considering the bipartition that divides the system into two distinguishable subsystems.
	
	The paper is organized as follows. In Sec. \ref{sec_entanglement_fermions} we give an overview of the notion and quantification of entanglement for two identical fermions, then, in Sec. \ref{sec_entanglement}, we propose the measure of entanglement for the hybrid system. In Sec. \ref{sec_wavefunction} we derive the expression for the eigenfunctions of the system, with a particular focus on the ground state. Section \ref{sect_rho_calculation} presents the procedure to obtain the reduced density matrices which are needed in order to compute the entanglement measure. We present and discuss our results regarding spatial correlations and entanglement in Sec. \ref{sec_results}. Finally, a summary and conclusions are given in Sec. \ref{sec_conclusions}.
	
	\section{Identical-fermions entanglement}
	\label{sec_entanglement_fermions}
	
	Before introducing the notion of fermionic entanglement measures it is convenient to review the basics of the definition and quantification of entanglement in distinguishable-parties systems. 
		We therefore start by considering a quantum system $S$ constituted by two distinguishable subsystems, $S_1$ and $S_2$. 
		Each subsystem is described in the Hilbert space $H_1$ and $H_2$, respectively, while the Hilbert space of the composite system is given by the tensor product of $H_1$ and $H_2$. 
		
		A pure state $\ket{\psi}$ in $H_S = H_1 \otimes H_2$ is said to be separable if and only if it can be written as a product of states belonging to each Hilbert subspace, so $\ket{\psi}_{\textrm{sep}} = \ket{\psi_1} \otimes \ket{\psi_2}$, while an entangled state is a non-separable one. As a consequence, in an entangled state one cannot assign a pure state to each subsystem, and therefore the states of $S_1$ and $S_2$ are necessarily mixed. The amount of entanglement of the pure global state of  $S=S_1+S_2$ can be therefore quantified by the degree of mixedness of the reduced density matrices $\rho_1$ and $\rho_2$.
		A widely used measure of entanglement is the von Neumann entropy of any of the marginal density matrices $\varrho=\rho_{1(2)}$, namely $S_{vN}(\varrho)= - \text{Tr}\, ( \varrho \ln \varrho )$, which by means of the Mercator series can be approximated by the linear entropy
		\beq\label{SLdis}
		S_L(\varrho) = 1- \text{Tr}(\varrho^2),
		\eeq
		with the advantage of being easier to compute \cite{horodecki_2009,nielsen_chuang_book}.

		When the constituents of $S$ are two indistinguishable fermions,  the anti-symmetrization postulate reduces the accessible Hilbert space to the antisymmetric subspace of the tensor product of single-particle Hilbert spaces. 
		Consequently, the composite system cannot occupy a factorizable state of the form $|\varphi\rangle\ \otimes|\phi\rangle$ (with $|\varphi\rangle$ and $|\phi\rangle$ orthogonal and normalized single-fermion state vectors), but instead occupies the following state, referred to as a Slater determinant:
		\begin{equation}
			\label{eq_one_slater}
			\ket{\psi^{sl}}=\frac{1}{\sqrt{2}}(\ket{\varphi}\otimes\ket{\phi}-\ket{\phi}\otimes\ket{\varphi}),   
		\end{equation}
		obtained by antisymmetrizing  the product $|\varphi\rangle\ \otimes|\phi\rangle$.
		
		The state (\ref{eq_one_slater}) introduces the so-called exchange correlations between fermions, which are inherent correlations that ensue exclusively due to the antisymmetrization requirement, and prevent the pair of parties from behaving independently from each other. This already points to the need of a redefinition of entanglement ---and consequently of the entanglement measures in identical-fermion systems--- in order to account solely for the system's correlations on top of the exchange correlations \cite{plastino_2009_epl, manzano_2010, lopez_rosa_2015, tichy_2011_JPB}. 
		In this context, fermionic entanglement is associated to the quantum correlations exhibited by the state beyond the minimal correlations originated from the anti-symmetry of the fermionic wave-function \cite{cuestas_2020_ferm}. 
		In line with this definition, a pure state of a composite of two indistinguishable fermions is considered non-entangled (or separable) if and only if it can be represented by a single Slater determinant, i.e., by a state of the form (\ref{eq_one_slater}).
		Slater determinants are thus the fermionic analogue of product states in systems composed of distinguishable particles: both states posses the minimal possible amount of correlations; in the distinguishable-party case such minimal value is zero, whereas in the fermionic case it encodes exchange correlations only.

		Whereas according to the present definition  separable fermionic states are not devoid of correlations,
		the notion of `separability' attributed to a single Slater determinant is consistent with the possibility of completely assign a set of properties to each particle \cite{ghirardi_2004,tichy_2011_JPB}.
		Indeed, if a composite of two fermions occupies the state (\ref{eq_one_slater}) then, with certainty, one of the particles is in the state $\ket{\varphi}$ while the other one is in the state $\ket{\phi}$. As a result, in a Slater determinant a complete set of commuting observables can be ascribed to the individuals, a feature that holds also for non-entangled (product) states in composites of distinguishable parties \cite{Majtey2023}.

	A natural generalization of the entanglement measure $S_L(\varrho)$ between two distinguishable parties leads to the following fermionic counterpart,  
	\eq
	\label{eq_SLf}
	S_F (\varrho)= 1- 2\, \text{Tr}(\varrho^2) \,
	\en
	where $\varrho$ denotes the reduced density matrix of a single fermion \cite{schielmann_2001,pauskaukas_2001}. The factor $2$ arises in order to exclude the exchange correlations, ensuring that $S_F$ vanishes whenever the complete state is a Slater determinant. 
	
	\section{Entanglement in a many-particle hybrid system}
	\label{sec_entanglement}
	
	Let us consider a system made up of fermions of two distinct species, namely $a$ and $b$. 
	We will denote the number of fermions of  species $a$ by $N_a$, and the number of fermions of species $b$ by $N_b$.
	The complete system is assumed to be in a pure state $|\psi\rangle$, which  complies with the antisymmetry requirements under the exchange of any two particles of the same species.
	
	As discussed above, when dealing with pure states of bipartite systems $S=S_1+S_2$, it is natural to use as an indicator of the amount of entanglement between $S_1$ and $S_2$ a measure based on the degree of mixedness of the reduced density matrices (the entropy $S_L$ for distinguishable parties, or the entropy $S_F$ for indistinguishable fermions). 
	In what follows we will generalize these purity-based entanglement measures to be of applicability in the hybrid system, in which fermions of the species $a$, are distinguishable from the fermions of type $b$, but are indistinguishable among themselves. 
	We start by considering a separable, minimally correlated state of the complete hybrid system, i.e., the tensor product of two Slater determinants, each one involving all the particles of the same species:
	\begin{equation}\label{eq_mix_sep}
		\ket{\psi}_{\textrm{sep}}=\ket{\psi^{sl}_{N_a}} \otimes \ket{\psi^{sl}_{N_b}}.
	\end{equation}
	Clearly this state has no correlations between the (distinguishable) $a$-type and $b$-type fermions, neither exhibits fermionic correlations among particles of the same species.
	
	Let us now consider the bipartition $(M_a+M_b)|(N_a-M_a+N_b-M_b)$, which divides the system into $M_a$ particles of kind $a$ plus $M_b$ particles of kind $b$, and the rest. The reduced state of the $M_a+M_b$ subsystem, namely $\rho_{M_a+M_b}$, is obtained by tracing out  $\rho = \ket{\psi}_{\text{sep}} \bra{\psi}_{\text{sep}} = \rho_{N_a}^{sl}\otimes \rho_{N_b}^{sl}$ (with $\rho_{N_a}^{sl}= |\psi^{sl}_{N_a}\rangle \langle\psi^{sl}_{N_a}|$, and similarly for $b$) over $N_a-M_a$ fermions of type $a$ and $N_b-M_b$ fermions of type $b$,
	\begin{eqnarray}
		\rho_{M_a+M_b}&=&\textrm{Tr}_{N_a-M_a;N_b-M_b}\big(\rho_{N_a}^{sl}\otimes \rho_{N_b}^{sl}\big)\nonumber\\
		&=&\textrm{Tr}_{N_a-M_a}\big(\rho_{N_a}^{sl}\big)\otimes \textrm{Tr}_{N_b-M_b}\big(\rho_{N_b}^{sl}\big)\nonumber\\
		&=&\rho_{M_a}\otimes \rho_{M_b}.
	\end{eqnarray}
	The purity of this reduced density matrix reads
	\begin{equation}
		\begin{split}
			\text{Tr}(\rho_{M_a+M_b}^2)& = \text{Tr}(\rho_{M_a}^2 \otimes \rho_{M_b}^2) \\
			&= \text{Tr}(\rho_{M_a}^2) \text{Tr}(\rho_{M_b}^2) \\
			&=\binom{N_a}{M_a}^{-1}\binom{N_b}{M_b}^{-1},
		\end{split}
	\end{equation}
	where in the last equality we used that for a pure state $\ket{\psi_N}$ of $N$ identical fermions, the purity of the (reduced) state of $M<N$ fermions is bounded as $\text{Tr}(\rho_{M}^2)\leq \binom{N}{M}^{-1}$, with the equality holding if and only if $\ket{\psi_N}$ is a Slater determinant \cite{majtey_2016, plastino_2009_epl}. We have thus shown that
	\beq
	\ket{\psi}_{\textrm{sep}}\Rightarrow\text{Tr}(\rho_{M_a+M_b}^2)=\binom{N_a}{M_a}^{-1}\binom{N_b}{M_b}^{-1}.
	\eeq
	
	Taking into consideration that for a minimally correlated state the purity of its reduced density matrices is maximal (there is maximal information about all the possible subsystems), the above result implies that the inequality
	\beq
	\text{Tr}(\rho_{M_a+M_b}^2) < \binom{N_a}{M_a}^{-1}\binom{N_b}{M_b}^{-1},
	\eeq
	is a signature of correlations across the bipartition $(M_a+M_b)|(N_a-M_a+N_b-M_b)$. This entanglement criterion leads us to consider 
	\begin{equation}
		\label{eq_e_bip_mix_Npairs}
		\varepsilon_{(M_a+M_b)|(N_a-M_a+N_b-M_b)} = 1-\binom{N_a}{M_a} \binom{N_b}{M_b}\text{Tr}(\rho_{M_a+M_b}^2)
	\end{equation}
	as an appropriate quantifier of entanglement in the hybrid composite system. For $M_a=N_a$ and $M_b=0$, meaning that the bipartition divides the system into two distinguishable subsystems, the measure reduces to the linear entropy (\ref{SLdis}). For any other case the measure $\varepsilon$  includes the binomial factors in order to exclude the exchange correlations among fermions of the same species. 
	In particular, it is straightforward to check the consistency of the measure proposed in Eq. \eqref{eq_e_bip_mix_Npairs} with that used in Refs. \cite{majtey_2016, plastino_2009_epl} to quantify the entanglement in a pure state of an arbitrary number of indistinguishable fermions. 
	
	It is worth mentioning that the distinguishable nature of fermions of different species is central for the validity of the entanglement measure in Eq. \eqref{eq_e_bip_mix_Npairs}. In particular, by construction, such measure is consistent with a separable state of the form \eqref{eq_mix_sep}, implying that the $N_a$ parties are distinguishable for the remaining $N_b$, and involving exchange correlations only between particles of the same species. Thus, for example, the proposed measure is unsuitable for a system of $N_a+N_b$ identical $1/2$-spin fermions  in which the spin were considered an actual degree of freedom (in addition to translational ones); in such case exchange correlations would exist among all the $N_a+N_b$ particles, and the appropriate (single-species) entanglement measure would be that advanced in  \cite{majtey_2016}. If, however, the spin projection of each particle is fixed, so its specific value defines the $a$ and $b$ type of particles, then Eq. \eqref{eq_e_bip_mix_Npairs}  fits well as a suitable measure of (multi-species) entanglement.
	
	In order to determine the entanglement (\ref{eq_e_bip_mix_Npairs}), the purity of the reduced density matrix of interest must be computed. 
	We will now revisit how this computation may be performed in terms of the matrix elements $\varrho(q,q')=\langle q|\varrho|q'\rangle$ of a generic reduced density operator $\varrho$ in an orthonormal continuous basis $\{\ket{q}\}$. In terms of these matrix elements, the purity of the state $\varrho$ reads
	\beq\label{rhopurity1}
	\textrm{Tr}(\varrho^2)=\int \langle q|\varrho^2|q\rangle\,dq=\int \langle q|\varrho|q'\rangle \langle q'|\varrho|q\rangle\,dq \,dq',
	\eeq
	where we used the completeness relation $\int \ket{q'}\bra{q'} dq'=\mathbb I.$ Due to the hermiticity of $\varrho$, and assuming the matrix elements to be real (which is the case for the scenario considered below) it holds that $\langle q'|\varrho|q\rangle=(\langle q|\varrho|q'\rangle)^*=\langle q|\varrho|q'\rangle$, whence
	\begin{eqnarray}\label{rhopurity2}
		\textrm{Tr}(\varrho^2)&=&\int \langle q|\varrho|q'\rangle \langle q|\varrho|q'\rangle\,dq\, dq'\\\nonumber
		&=&\int \varrho^2(q,q')dq\, dq'.
	\end{eqnarray}
	We will come back to this expression in Section \ref{sect_rho_calculation}. 
	
	\section{Balanced two-component harmonic system }
	\label{sec_wavefunction}
	
	In what follows we consider a system made up of $2N$ fermions, $N$ of which are of type $a$ and $N$ are of kind $b$.
	In this section we will focus on the specific dynamics of the hybrid system, and solve the appropriate ground-state wave function considering the Moshinsky model, as shown below.
	
	In the following we assign the coordinates $x_1,...,x_N$ to particles of kind $a$, and $x_{N+1},...,x_{2N}$ to particles of kind $b$, and focus on the $N$-pair  Hamiltonian
	\begin{equation}
		\mathcal{H} =  \frac{1}{2}\sum_{i=1}^{2N}\left(-\frac{\hbar^2}{m}\frac{\partial^2}{\partial x_i^2} + m \omega^2 x_i^2\right)  - \Lambda \frac{ m \omega^2}{2} \sum_{i=1}^{N}\sum_{j=N+1}^{2N}\left(x_i-x_j\right)^2,
	\end{equation}
	where $m$ stands for the mass of each particle, $\omega$ for the oscillation frequency and $\Lambda$ for a real coupling constant, whose sign distinguishes the attractive regime from the repulsive one [see below Eq. (\ref{Hsep})]. 
	
	In terms of the dimensionless variables  $x_i \rightarrow \sqrt{\frac{m \omega}{\hbar}} x_i$ and $\mathcal{H} \rightarrow \frac{\mathcal{H}}{\hbar \omega}$ the Hamiltonian reduces to
	\begin{equation}
		\mathcal{H} = \frac{1}{2}\sum_{i=1}^{2N}\left(-\frac{\partial^2}{\partial x_i^2} + x_i^2\right) - \frac{ \Lambda}{2} \sum_{i=1}^{N}\sum_{j=N+1}^{2N}\left(x_i-x_j\right)^2.
	\end{equation}
	We now rewrite the last term noticing that
	\begin{eqnarray}
		\sum_{i=1}^{N}\sum_{j=N+1}^{2N}\left(x_i-x_j\right)^2
		&=&N \sum_{i=1}^{2N} x_i^2 -2 \sum_{i=1}^{N}\sum_{j=N+1}^{2N} x_i x_j\\
		&=&N \sum_{i=1}^{2N} x_i^2 - \sum_{i=1}^{N}\sum_{j=N+1}^{2N} \left( x_i x_j + x_j x_i\right) \\
		&=&N \sum_{i=1}^{2N} x_i^2 - \sum_{i,j=1}^{2N}x_i \mathcal{K}_{ij}x_j,
	\end{eqnarray}
	where $\mathcal{K}_{ij}$ stands for the $(i,j)$ element of the $2N\times 2N$ matrix $\mathcal{K}$, given by 
	\begin{equation}
		\mathcal{K}=\left(
		\begin{array}{ c c c c c c }
			0 & \cdots  & 0 & 1 & \cdots  & 1\\
			\vdots  & \ddots  & \vdots  & \vdots  & \ddots  & \vdots \\
			0 & \cdots  & 0 & 1 & \cdots  & 1\\
			1 & \cdots  & 1 & 0 & \cdots  & 0\\
			\vdots  & \ddots  & \vdots  & \vdots  & \ddots  & \vdots \\
			1 & \cdots  & 1 & 0 & \cdots  & 0
		\end{array}\right).
	\end{equation}
	Then, we can write the Hamiltonian as
	\begin{equation}\label{Hred}
		\mathcal{H} = \frac{1}{2}\sum_{i=1}^{2N}\left[-\frac{\partial^2}{\partial x_i^2} + (1-N\Lambda)x_i^2\right] + \frac{ \Lambda}{2}\sum_{i,j=1}^{2N}x_i \mathcal{K}_{ij}x_j.
	\end{equation}
	
	In order to find the normal modes of $\mathcal H$, we focus on the eigenvectors of $\mathcal{K}$. This matrix has two non-zero eigenvalues, namely $N$ and $-N$. The normalized eigenvector corresponding to the eigenvalue $N$ is
	\beq
	v^{(1)}=\frac{1}{\sqrt{2N}}(1, \dots, 1)^T,
	\label{eq:CM-tot}
	\eeq
	and the normalized eigenvector with eigenvalue $-N$ reads
	\beq
	v^{(2N)}=\frac{1}{\sqrt{2N}}(1, \dots, 1, -1, \dots, -1)^T.
	\label{eq:rel-AB}
	\eeq
	The null eigenvalue has multiplicity equal to $2N-2$ (since $\mathcal{K}$ has only two linearly independent rows). 
	We will denote with $\{v^{(2)},...,v^{(2N-1)}\}$ some orthonormal eigenbasis of the subspace corresponding to the eigenvalue zero. With this aim we introduce an orthogonal transformation between the variables $\{x_i\}$ and a new set of variables $\{R_i\}$, such that
	\begin{equation}
		R_i=\sum_{j=1}^{2N}v^{(i)}_j x_j, \quad i=1,...,2N .
		\label{eq:change_of_var}
	\end{equation}
	In particular, from Eq. \eqref{eq:CM-tot} we get
	\begin{equation}
		R_1=\frac{1}{\sqrt{2N}}\sum^{2N}_{i=1}x_i = \sqrt{2N} R,
		\label{eq:total_cm_in_x}
	\end{equation}
	where $R$ denotes the total center of mass, while Eq. \eqref{eq:rel-AB} gives
	\begin{equation}
		R_{2N}=\frac{1}{\sqrt{2N}}\Big(\sum^{N}_{i=1}x_i-\sum^{2N}_{j=N+1}x_j\Big) = \sqrt{\frac{N}{2}} (R_a-R_b) ,
		\label{eq:cm_AB_in_x}
	\end{equation}
	with $R_{a(b)}$ being the center of mass of the particles of species $a(b)$. It is worth mentioning that the new variables defined in Eq. \eqref{eq:change_of_var} are similar to the Jacobi variables used to solve the many-body system of distinguishable \cite{koscik_2013} or indistinguishable \cite{benavides2014entanglement} particles interacting via a harmonic potential. However, an important difference arises because we are considering particles of different types. 
	Our transformation is applied to the subspace of particles of each kind in the same way as for the single species in Refs. \cite{koscik_2013, benavides2014entanglement}, and it also defines the total center of mass and the distance between the center of mass of each type of particles, as is explicit from  Eqs. \eqref{eq:total_cm_in_x} and \eqref{eq:cm_AB_in_x}, respectively.
	
	Resorting to the variables $\{R_i\}$ we transform the Hamiltonian (\ref{Hred}) into a separable form: 
	\begin{equation}\label{Hsep}
		\mathcal{H}=\frac{1}{2}\left(-\frac{\partial^2}{\partial R_1^2} + R_1^2\right) + \frac{1}{2}\left(-\frac{\partial^2}{\partial R_{2N}^2} + \kappa^2_2R_{2N}^2\right)
		+\frac{1}{2}\sum_{i=2}^{2N-1}\left(-\frac{\partial^2}{\partial R_i^2}+\kappa^2_1R_{i}^2\right),
	\end{equation}
	where we defined $\kappa_1=\sqrt{1-N\Lambda}$ and $\kappa_2=\sqrt{1-2N\Lambda}$. The attractive and repulsive regimes are associated to the conditions $\Lambda < 0$ and $0 < \Lambda \leq 1/(2N) $, respectively, leading in both cases to an effective real and positive frequency that translates into the existence of a bound state \cite{Glasser_2013,benavides2014entanglement}. 
	When $2N \Lambda > 1$, or equivalently, $2N (m\Lambda\omega^2/2) > m\omega^2/2$, i.e. when the magnitude of the interparticle coupling constant multiplied by the total number of particles exceeds the confinement strength, the system does not have a bound state. 
	
	Equation (\ref{Hsep}) can be identified with the Hamiltonian of a set of $2N$ uncoupled harmonic oscillators. Recalling that the eigenfunctions $\varphi_n(x)$ of the harmonic oscilator satisfy
	\begin{equation}
		\frac{1}{2}\left(-\frac{\partial^2}{\partial x^2}+x^2\right)\varphi_n(x)=\left(n+\frac{1}{2}\right)\varphi_n(x),
	\end{equation}
	so for $\kappa >0$,
	\begin{equation}
		\frac{1}{2}\left(-\frac{\partial^2}{\partial x^2}+\kappa^2 x^2\right)\varphi_n(\sqrt{\kappa}x)=\kappa \left(n+\frac{1}{2}\right)\varphi_n(\sqrt{\kappa}x),
	\end{equation}
	the eigenfunctions of the many-body Hamiltonian of Eq. \eqref{Hsep} can be written as
	\begin{equation}\label{Hsol}
		\psi =  \varphi_{n_1}\left(R_1\right)\kappa_2^{1/4}\varphi_{n_{2N}}\left(\sqrt{\kappa_2}R_{2N}\right)
		\kappa_1^{(N-1)/2}\varphi_{n_2}\left(\sqrt{\kappa_1}R_2\right)...\varphi_{n_{2N-1}}\left(\sqrt{\kappa_1}R_{2N-1}\right),
	\end{equation}
	where we wrote $\kappa^{1/4}\varphi_n(\sqrt{\kappa}x)$ in order to guarantee the normalization of the eigenstates. 
	The associated energy levels read
	\begin{equation}
		\label{eq_energy}
		\frac{E}{\hbar \omega}= n_1 + \frac{1}{2} + \kappa_1 ( n_2 + \cdots + n_{2N-1} + N-1) + \kappa_2 \big(n_{2N}+\frac{1}{2}\big) ,
	\end{equation}
	where the quantum numbers $n_i=0,1,\dots$ (with $i=1,\dots,2N$) represent excitations of oscillators in the new set of variables $\{R_i\}$, or normal modes.
	
	\subsection{Antisymmetrization of the eigenfunctions}
	
	We will now construct the appropriate combination of solutions \eqref{Hsol}, in order to comply with the antisymmetry requirement under the exchange of any pair of particles of the same kind. To this aim we first observe that the coordinates $R_1$ and $R_{2N}$ are invariant under the exchange of two particles of the same species [see Eq. \eqref{eq:total_cm_in_x} and \eqref{eq:cm_AB_in_x}]. As for the other $2N-2$ coordinates, namely $R_2,\dots,R_{2N-1}$, related to the eigenvectors of $\mathcal K$ with null eigenvalue, it is convenient to avoid mixing the variables associated with different kind of particles. 
	This can be done by taking the $v^{(i)}$ vectors as
	\begin{equation}\label{vi1}
		v^{(i)}=(v^{(i)}_1, \dots, v^{(i)}_{N},\, 0, \dots, 0)^T, \quad i=2,...,N,
	\end{equation}
	with $\sum_{j=1}^{N}v^{(i)}_j=0$, and
	\begin{equation}
		v^{(i)}=(0, \dots, 0, \, v^{(i)}_{N+1}, \dots, v^{(i)}_{2N})^T, \quad i=N+1,...,2N-1,
	\end{equation}
	with $\sum_{j=N+1}^{2N}v^{(i)}_j=0$. In particular, we choose the following elements of the vector (\ref{vi1}) for $i=2,\dots,N$, and $j=1,\dots,N$, 
	\begin{equation}
		v^{(i)}_j=\frac{1}{\sqrt{i(i-1)}}\times
		\begin{cases}
			-1  &  j<i, \\
			i-1  &  j=i,  \\
			0 & j>i.
		\end{cases}
		\label{eq:basis_elements}
	\end{equation}
	The remaining $N-1$ vectors are defined as  $v^{(N-1+i)}_{j+N}=v_{j}^{(i)}$ for $i=2,\dots,N$ and $j=1,\dots,N$ (note that, as expected, this is an orthonormal basis). In this way $R_2,\dots,R_{N}$ involve only  coordinates of $a$-type particles, whereas  $R_{N+1},\dots,R_{2N-1}$ involve only  coordinates of $b$-type particles. Consequently, when applying the antisymmetric operator  $\mathcal A_A^{\{x_i\}}$ (that antisymmetrizes with respect to the $a$-type fermions in the original spatial variables $\{x_i\}$) to the factor $\varphi_{n_2}\cdots\varphi_{n_{2N-1}}$ of the state (\ref{Hsol}), only the coordinates $R_2,\dots,R_{N}$ will be affected. Analogously, application of the antisymmetric operator $\mathcal A_B^{\{x_i\}}$ will only affect the coordinates $R_{N+1},\dots,R_{2N-1}$.  
	This implies that the antisymmetric wave function that ensues from (\ref{Hsol}) has the structure 
	\begin{equation}
		\label{solgen}
		\begin{split}
			\psi \propto \, & \varphi_{n_1}\left(R_1\right)\varphi_{n_{2N}}\left(\sqrt{\kappa_2}R_{2N}\right)\\
			&\times\mathcal{A}_A^{\{x_i\}}\Big[\varphi_{n_2}\left(\sqrt{\kappa_1}R_2\right)\cdots\varphi_{n_{N}}\left(\sqrt{\kappa_1}R_{N}\right)\Big]\\
			&\times\mathcal{A}_B^{\{x_i\}}\Big[\varphi_{n_{N+1}}\left(\sqrt{\kappa_1}R_{N+1}\right)\cdots\varphi_{n_{2N-1}}\left(\sqrt{\kappa_1}R_{2N-1}\right)\Big],
		\end{split}
	\end{equation}
	where
	\begin{equation}
		\label{antipsi}
		\begin{split}
			&\mathcal{A}_A^{\{x_i\}} \phi(x_1,\dots,x_N,x_{N+1},\dots,x_{2N})\\
			&=\sum_{\alpha_1,\dots,\alpha_N} \epsilon^{\alpha_1,\dots,\alpha_N} \phi(x_{\alpha_1},\dots,x_{\alpha_N},x_{N+1},\dots,x_{2N}),
		\end{split}
	\end{equation}
	and
	\begin{equation}
		\begin{split}
			&\mathcal{A}_B^{\{x_i\}} \phi(x_1,\dots,x_N,x_{N+1},\dots,x_{2N})\\
			&=\sum_{\alpha_{N+1},\dots,\alpha_{2N}} \epsilon^{\alpha_{N+1},\dots,\alpha_{2N}} \phi(x_1,\dots,x_N,x_{\alpha_{N+1}},\dots,x_{\alpha_{2N}}),
		\end{split}
	\end{equation}
	with $\epsilon^{\alpha_i,\alpha_j\dots}$ the antisymmetric Levi-Civita symbol. Note that since some terms can be `turned-off' by the antisymetrization operator, the norm of the resulting state can be less than the one of the input state, i.e., the antisymmetrization operation does not preserve the norm of the state. 
	
	We now observe that
	\begin{equation}
		\begin{split}
			&\varphi_{n_2}\left(\sqrt{\kappa_1}R_2\right)\cdots\varphi_{n_{2N-1}}\left(\sqrt{\kappa_1}R_{2N-1}\right)\\
			&\propto e^{-\frac{\kappa_1}{2}\sum_{i=2}^{2N-1}R_i^2} H_{n_2}\left(\sqrt{\kappa_1}R_2\right)\cdots H_{n_{2N-1}}\left(\sqrt{\kappa_1}R_{2N-1}\right),
		\end{split}
	\end{equation}
	with $H_m$ being the $m$-th Hermite polynomial. As $\lbrace R_i\rbrace$ and $\lbrace x_i\rbrace$ are related via an orthogonal transformation that preserves the inner product, $\sum_{i=1}^{2N}R_i^2=\sum_{i=1}^{2N}x_i^2$, then the expression in the exponential takes the form,
	\begin{eqnarray}
		\sum_{i=2}^{2N-1}R_i^2&=&\sum_{i=1}^{2N}x_i^2-R_1^2-R_{2N}^2 \nonumber\\
		&=&\sum_{i=1}^{2N}x_i^2-\frac{1}{2N}\left[\left(\sum_{i=1}^{2N}x_i\right)^2+\left(\sum_{i=1}^{N}x_i - \sum_{i=N+1}^{2N}x_i\right)^2\right]\nonumber\\
		&=&\frac{1}{N}\left[\sum_{1 \leq i < j\leq N}(x_i-x_j)^2 + \sum_{N+1 \leq i < j\leq 2N}(x_i-x_j)^2 \right],
		\label{eq:sum_sq_exp}
	\end{eqnarray}
	which is invariant under the exchange of particles of the same kind. Therefore, the antisymmetrization in (\ref{antipsi}) is carried out only over the Hermite polynomials:
	\begin{equation}
		\label{eq_as_inA}
		\mathcal{A}_A^{\{x_i\}} \Big[H_{n_2}\left(\sqrt{\kappa_1}R_2\right)\cdots H_{n_{N}}\left(\sqrt{\kappa_1}R_{N}\right)\Big],
	\end{equation}
	and
	\begin{equation}
		\label{eq_as_inB}
		\mathcal{A}_B^{\{x_i\}} \Big[H_{n_{N+1}}\left(\sqrt{\kappa_1}R_{N+1}\right) \cdots H_{n_{2N-1}}\left(\sqrt{\kappa_1}R_{2N-1}\right)\Big].
	\end{equation}
	
	\subsection{Ground state eigenfunction}  
	
	In order to construct the ground state we need to find the set of quantum numbers $\{n_1,\dots,n_{2N}\}$ that minimize the energy given in Eq. \eqref{eq_energy}. 
	Because $R_1$ and $R_{2N}$ are symmetric under the exchange of particles of type $a$ and $b$, in order to have the minimal energy we can set $n_1$ and $n_{2N}$ equal to $0$. Then the problem reduces to finding the minimum of the sum $n_2 + \dots + n_{2N-1}$, where $n_2, \dots, n_N$ are quantum numbers purely related to particles of kind $a$, whereas $n_{N+1}, \dots, n_{2N-1}$ are quantum numbers purely related to particles of kind $b$. 
	As proved in Ref. \cite{procesi2007lie}, a polynomial $P^{(A)}(u_1,\dots,u_N)$ is antisymmetric in its variables $u_1,\dots,u_N$ if and only if it is of the form $P^{(A)}(u_1,\dots,u_N) = V(u_1,\dots,u_N) \times P^{(S)}(u_1,\dots,u_N)$, where $P^{(S)}(u_1,\dots,u_N)$ denotes a symmetric polynomial and $V(u_1,\dots,u_N)=\prod_{i<j} (u_i-u_j)$ is the Vandermonde determinant. Both Eq. \eqref{eq_as_inA} and \eqref{eq_as_inB} are antisymmetric polynomials in terms of the ${x_i}$ variables, therefore we have
	\begin{equation}
		\label{eq_as_inA_2}
		\mathcal{A}_A^{\{x_i\}} \Big[H_{n_2}\left(\sqrt{\kappa_1}R_2\right)\cdots H_{n_{N}}\left(\sqrt{\kappa_1}R_{N}\right)\Big] = \prod_{1 \leq i<j \leq N} (x_i-x_j) P_A^{(S)}(x_1,\dots,x_N) ,
	\end{equation}
	and
	\begin{equation}
		\label{eq_as_inB_2}
		\begin{split}
			& \mathcal{A}_B^{\{x_i\}} \Big[H_{n_{N+1}}\left(\sqrt{\kappa_1}R_{N+1}\right)\cdots H_{n_{2N-1}}\left(\sqrt{\kappa_1}R_{2N-1}\right)\Big]= \\
			& \prod_{N+1 \leq i<j \leq 2N} (x_i-x_j) \times P_B^{(S)}(x_{N+1},\dots,x_{2N}) .
		\end{split}
	\end{equation}
	The degree of the polynomial in the $\{R_i\}$ variables on the right side of these two expression is $n_2+\dots+ n_N$ for Eq. \eqref{eq_as_inA_2}, and $n_{N+1}+\dots+n_{2N-1}$ for Eq. \eqref{eq_as_inB_2}.
	Since the variables $\{R_i\}$ are linear combinations of the $\{x_i\}$, the degree is the same in the $\{x_i\}$ variables. 
	Further, since the order of the Vandermonde determinant is $N(N-1)/2$, the order of the symmetric polynomial $P_A^{(S)}(x_1,\dots,x_N)$ is $n_2+\dots+n_N - N(N-1)/2 \geq 0$ and, equivalently, the order of $P_B^{(S)}(x_{N+1},\dots,x_{2N})$ is $n_{N+1}+ \dots +n_{2N-1} - N(N-1)/2 \geq 0$. 
	All this means that $n_2+\dots+n_N \geq N(N-1)/2 $ and $n_{N+1}+ \dots +n_{2N-1} \geq N(N-1)/2$. The quantum numbers that determine the ground state are therefore those that make the order of both polynomials $P_A^{(S)}$ and $P_B^{(S)}$ equal to zero, thus leading to 
	\begin{subequations}
		\begin{equation}
			\label{eq_as_inA_gs_R}
			\mathcal{A}_A^{\{x_i\}} \Big[ H_{n_2}\left(R_2\right)\cdots H_{n_N}\left(R_N\right)\Big] \propto \prod_{1 \leq i<j \leq N} (x_i-x_j)  ,
		\end{equation}
		with 
		\beq 
		\label{sumn1}
		n_2+\dots+n_N = \frac{N(N-1)}{2},
		\eeq
	\end{subequations}
	and
	\begin{subequations}
		\begin{equation}
			\label{eq_as_inB_gs_R}
			\mathcal{A}_B^{\{x_i\}}  \Big[ H_{n_{N+1}}\left(R_{N+1}\right)\cdots H_{n_{2N-1}}\left(R_{2N-1}\right)\Big] \propto  \prod_{N+1 \leq i<j \leq 2N} \left(x_i-x_j\right)   
		\end{equation}
		provided 
		\beq
		\label{sumn2}
		n_{N+1}+\dots+n_{2N-1} = \frac{N(N-1)}{2}.
		\eeq
	\end{subequations}
	
	An arbitrary choice of quantum numbers $\{n_2,\dots,n_{N}\}$ satisfying the condition (\ref{sumn1}) will lead either to Eq. (\ref{eq_as_inA_gs_R}), or to $\mathcal{A}_A^{\{x_i\}} \Big[ H_{n_2}\left(R_2\right)\cdots H_{n_N}\left(R_N\right)\Big] =0$. In other words, there always exists a set of quantum numbers $\{n_2,\dots,n_{N}\}$ that yield to a Vandermonde determinant. The same argument holds for $\{n_{N+1},\dots,n_{2N-1}\}$ under the condition \eqref{sumn2}. 
	
	The Vandermonde determinant has been shown to satisfy \cite{lando_2004_book,osenda_2015},
	\begin{equation}
		\prod_{1 \leq i<j \leq N} \left(u_i-u_j\right)  = 2^{-\frac{N(N-1)}{2}} \mathcal{A}^{\{u_i\}}_A\Big[ H_{0}\left(u_1\right)\cdots H_{N-1}\left(u_N\right)\Big],
	\end{equation}
	therefore, the two determinants in the ground state are given by
	\begin{equation}
		\label{eq_as_inA_gs_x}
		\prod_{1 \leq i<j \leq N} (x_i-x_j) = 2^{-\frac{N(N-1)}{2}} \mathcal{A}_A^{\{x_i\}} \Big[ H_{0}\left(x_1\right)\cdots H_{N-1}\left(x_N\right)\Big] ,
	\end{equation}
	and
	\begin{equation}
		\label{eq_as_inB_gs_x}
		\prod_{N+1 \leq i<j \leq 2N}  \left(x_i-x_j\right) = 2^{-\frac{N(N-1)}{2}}   \mathcal{A}_B^{\{x_i\}}\Big[ H_{0}\left(x_{N+1}\right)\cdots H_{N-1}\left(x_{2N}\right)\Big] . 
	\end{equation}
	
	From Eqs. (\ref{eq_energy}), (\ref{sumn1}), and (\ref{sumn2}), and recalling that $\kappa_1=\sqrt{1-N\Lambda}$ and $\kappa_2=\sqrt{1-2N\Lambda}$, we conclude that the energy of the ground state is  
	\begin{equation}
		\label{eq_energyMIN}
		\frac{E_0}{\hbar \omega} = \sqrt{1-N\Lambda} (N^2-1) + \frac{\sqrt{1-2N\Lambda}+1}{2} ,
	\end{equation}
	which can be seen as the minimization of two sets of $N$ quantum numbers $n_i = 0,1,2,\dots$, where each set associated to particles of one kind is subject to the condition imposed by the Pauli principle among fermions of the same species (having all distinct numbers for each particle). 
	
	Notice that the energy of the ground state is an increasing function of the parameter $N\Lambda$ (which is proportional to the magnitude 
	$|\Lambda|$ of the interaction strength and the total number of particles). In the non-interacting case ($\Lambda = 0$) the ground state energy reads $E_0/\hbar \omega = N^2$, which is twice the energy of $N$ non-interacting fermionic oscillators. For a large number of particles ($N \gg 1$) and attractive interaction ($\Lambda < 0$) or in the strong attractive limit ($-\Lambda \gg 1$), Eq. \eqref{eq_energyMIN} can be rewritten as $E_0/(\hbar \omega \sqrt{-N\Lambda}) \sim  N^2 + (1/\sqrt{2})-1$. This shows that in the attractive regime for large $N$, or in the strongly attractive limit, the energy of the ground state equals the energy of a binary non-interacting mixture of fermions with a rescaled or effective frequency $\omega_{\text{eff}}= \omega \sqrt{-N\Lambda}$, minus a constant correction that lowers the total energy. 
	For a large repulsive interaction [$0 < \Lambda \sim 1/(2N)$] we can write $E_0/(\hbar \omega/\sqrt{2}) \sim N^2+ (1/\sqrt{2})-1$, i.e. the energy of the ground state is given by the energy of a binary non-interacting mixture of fermions with a new rescaled frequency $\omega_{\text{eff}}=\omega / \sqrt{2}$, minus the same constant correction obtained for the previous case equal to $1/\sqrt{2}-1 \sim -0.293$. 
	
	In the presence of a harmonic trap the volume has an inverse relation with the effective frequency of the harmonic oscillator \cite{romero_rochin_2005_prl,romero_rochin_2005_jpcb}. Then, we can estimate the fraction between the volume $V_S$ occupied by the system  and the volume $V_T$ due to the confinement or trap  as $\mathcal V = V_{S}/V_{T}$. From the previous reasoning we conclude that in the non-interacting case (in which $\omega_{\text{eff}}=\omega$) the system effectively fills all the available volume, so $\mathcal V = 1$; in the strongly repulsive limit the system natural scale competes with the one of the trap and $\mathcal V = \sqrt{2}>1$; finally, in the strongly attractive (or attractive with large $N$) regime the volume occupied by the system vanishes as $\mathcal V = 1/\sqrt{- N \Lambda}$. 
	
	To sum up, the ground state can be rewritten as        
	\begin{equation}
		\label{eq_solucion}
		\begin{split}
			\psi_0 = & \, \mathcal{N}_N e^{-\frac{1}{4N}\left(\sum\limits_{i=1}^{2N}x_i\right)^2} e^{-\frac{\sqrt{1-2N \Lambda}}{4N}\left(\sum\limits_{i=1}^{N}x_i - \sum\limits_{j=N+1}^{2N}x_j\right)^2}\\
			&\times  \prod_{1 \leq i<j \leq N} e^{-\frac{\sqrt{1-N \Lambda}}{2N}\left(x_i-x_j\right)^2}\left(x_i-x_j\right)\\
			& \times  \prod_{N+1 \leq i<j \leq 2N} e^{-\frac{\sqrt{1-N \Lambda}}{2N}\left(x_i-x_j\right)^2}\left(x_i-x_j\right),
		\end{split}
	\end{equation}
	where $\mathcal{N}_N$ is the normalization constant (see \ref{sec_app_normalization}),  
	\begin{equation}
		\mathcal{N}_N = \kappa_2^{1/4}\,  \kappa_1^{(N^2-1)/2}\, 2^{\frac{N(N-1)}{2}}\left[ N! \, \pi^{N/2}\, \prod_{k=0}^{N-1} k!\right]^{-1}.
	\end{equation}
	
	Before concluding this section we highlight that the part corresponding to the relative coordinates of the ground state of a system of $N$ pairs with interaction between particles of different species is equivalent to that of the ground state of a system of $N$ identical fermions, with all particles interacting with each other \cite{benavides2014entanglement} (see \ref{sec_app_solutions_N_fermions}). 
	To make this relation explicit we write
	\begin{equation}
		\label{eq_psi_in_ayb}
		\psi_0 = \kappa_2^{1/4} \psi_A(\{x_a\}) \psi_B(\{x_b\}) e^{-\frac{\sqrt{1-2N\Lambda}-1}{4} N (R_a-R_b)^2} ,
	\end{equation}
	where $\psi_{A(B)}({x_{a(b)}})$ denotes the state of a system of $N$ identical fermions of kind $a(b)$ and $R_{a(b)}$ is the center of mass of the $a(b)$ particles with coordinates $\{x_{a(b)}\} = x_{1(N+1)},\dots,x_{N(2N)} $. 
	In the above expression the limit $\Lambda \to 0$ can be straightforwardly checked, leading, as expected, to a state that is separable in the particles of kind $a$ and $b$. It is also apparent the exchange symmetry between particles of different kind, i.e. $ \{x_a\} \leftrightarrow \{x_b\}$, a symmetry of the Hamiltonian which is not explicit when resorting to the $\{R_i\}$ variables. Moreover, from Eq. \eqref{eq_psi_in_ayb} we conclude that the correlations between particles of different kind arise from a center of mass mode (the last factor is the ground state of an oscillator in the distance between the center of mass of particles of each species).
	
	\section{Procedure to compute reduced density matrices}
	\label{sect_rho_calculation}
	
	As discussed in Section \ref{sec_entanglement} (here with $N_a=N_b=N$), in order to quantify the entanglement across a generic bipartition  $(M_a+M_b)|(2N-M_a-M_b)$  in the hybrid composite system, it is necessary to compute the purity of the reduced density matrix $\rho_{M_a+M_b}$. To this aim, we will now present the method employed to compute such density matrices, from a given wave function $\psi$.
	
	In general, if we consider the bipartition in which  one of the subsystems has $M_a$ particles of type $a$ and $M_b$ of type $b$, we are interested in evaluating the matrix elements of the reduced density matrix
	\begin{equation}
		\rho_{M_a+M_b}=\int\langle \mathbb{X}_{2N-M_a-M_b}|\psi\rangle \langle\psi|\mathbb{X}_{2N-M_a-M_b}\rangle \, d\mathbb{X}_{2N-M_a-M_b},
	\end{equation}
	where $\mathbb{X}_{2N-M_a-M_b}$ stands for the set of coordinates of particles in the bipartition containing $N-M_a$ particles of type $a$ and $N-M_b$ particles of type $b$, so
	\beq
	\mathbb{X}_{2N-M_a-M_b}=\{x_{M_a+1},\dots,x_{N},\, x_{N+M_b+1},\dots,x_{2N}\}.
	\eeq
	Analogously, writing $\mathbb{X}_{M_a+M_b}$ for the set of coordinates 
	\beq
	\mathbb{X}_{M_a+M_b}=\{x_1,\dots,x_{M_a},\, x_{N+1},\dots,x_{N+M_b}\},
	\eeq
	we get for the matrix elements of  $\rho_{M_a+M_b}$ the following:
	\begin{equation}
		\begin{split}\label{int}
			&\rho_{M_a+M_b}\left(\mathbb{X}_{M_a+M_b}, \mathbb{X}_{M_a+M_b}^\prime\right)=\langle \mathbb{X}_{M_a+M_b}|\rho_{M_a+M_b}|\mathbb{X'}_{M_a+M_b}\rangle
			\\
			&= \int \psi\left(\mathbb{X}_{M_a+M_b},\mathbb{X}_{2N-M_a-M_b}\right)\psi\left(\mathbb{X}_{M_a+M_b}^\prime,\mathbb{X}_{2N-M_a-M_b} \right) d\mathbb{X}_{2N-M_a-M_b},
		\end{split}		
	\end{equation}
	where we assumed that the wave functions are real.
	
\begin{figure}[tb]
		\centering
		\includegraphics[width=0.75\columnwidth]{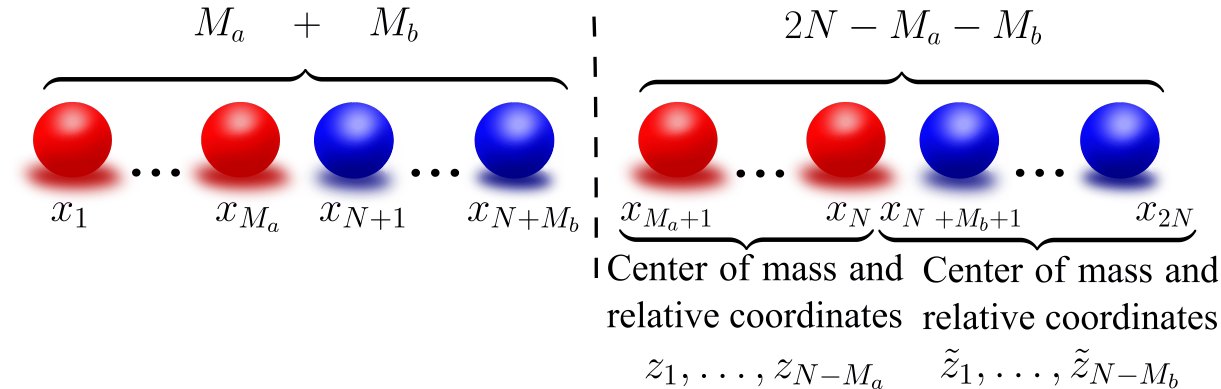}
		\caption{Change of variables used to compute the integral needed to get the reduced density matrix over a bipartition, here represented as the dashed vertical line. The red and blue circles are associated to particles of kind $a$ and $b$, respectively.}
		\label{fig:variable_change_rdm1}       
	\end{figure}

	Performing the integral (\ref{int}) in the actual coordinates of the particles can be extremely difficult, so we look for a change of variables that simplifies the integration. With this in mind we first consider a generic set of $n$ variables $\{r_1,\dots,r_n\}$ and define the corresponding `center of mass' ($z_1$) and `relative coordinates' ($z_i$, with $i=2,\dots,n$) as
	\begin{equation}
		\label{eq_transformation_2}
		\begin{split}
			&z_{1} =\frac{1}{\sqrt{n}}\sum_{j=1}^{n} r_{j}, \\
			&z_{i} =\frac{1}{\sqrt{i( i-1)}}\left[( i-1) r_{i} -\sum _{j=1}^{i-1} r_{j}\right].
		\end{split}
	\end{equation}
	The main idea is to identify a set $\{r_1,\dots,r_{N-M_a}\}$ with the set of coordinates of particles of kind $a$ in one of the bipartitions, and a set $\{r'_1,\dots,r'_{N-M_b}\}$ with the set of coordinates of particles of kind $b$ in that same bipartition. This procedure is depicted in Fig. \ref{fig:variable_change_rdm1}. We denote with $z_1, \dots, z_{N-M_a}$ the transformed coordinates in the first case, and $\tilde{z}_1, \dots, \tilde{z}_{N-M_b}$ the transformed coordinates in the second one, and perform the integrals over $\mathbb{X}_{2N-M_a-M_b}$ in terms of these variables (exploiting the fact that writing the wave function in these variables simplifies the terms in the exponential part). 
	
	\subsection{Bipartition $a$-type vs $b$-type particles}
	
	Let us resort to the aforementioned method to compute the (reduced) density matrix of the subsystem containing all $a$-type fermions [which corresponds to the bipartition $(M_a+M_b)|(2N-M_a-M_b)$ with $M_a=N$ and $M_b=0$]. In order to stress that the $N$ particles of the subsystem of interest are all of the same species $a$, we will denote the corresponding density matrix as $\rho_{N_a}$, and the bipartition as $N_a|N_b$. In this case, we must perform the transformation (\ref{eq_transformation_2}) to the coordinates of all the particles of kind $b$. After this change of variables (with $\tilde{z}$ denoting the new coordinates of $b$-type particles), the ground state wave function reads
	\begin{equation}
		\begin{split}
			\psi_0 = & \mathcal{N}_N e^{-\frac{1}{4N}\left(\sum_{i=1}^{N}x_i+\sqrt{N}\tilde{z}_1\right)^2} e^{-\frac{\kappa_2}{4N}\left(\sum_{i=1}^{N}x_i - \sqrt{N}\tilde{z}_1\right)^2}\\
			&\times \prod_{1 \leq i<j \leq N} e^{-\frac{\kappa_1}{2N}\left(x_i-x_j\right)^2}\left(x_i-x_j\right)\\
			& \times e^{-\frac{\kappa_1}{2}\sum_{i=1}^{N-1} \tilde{z}_i^{2}} \widetilde{V}(\tilde{z}_2, \dots, \tilde{z}_N),
		\end{split}
	\end{equation}
	where in order to simplify the notation we wrote
	\begin{equation}
		\widetilde{V}(\tilde{z}_2, \dots, \tilde{z}_N)=V[x_{N+1}(\tilde{z}_2, \dots, \tilde{z}_N),\dots,x_{2N}(\tilde{z}_2, \dots, \tilde{z}_N)] .
	\end{equation}
	
	As follows from Eq. (\ref{int}), the matrix elements of the reduced density matrix of all fermions of type $a$ is thus given by 
	\begin{equation}
		\begin{split}
			\rho_{N_a}(\mathbb{X}_{N_a},\mathbb{X}_{N_a}') & =  \int \psi_0(\mathbb{X}_{N_a},\mathbb{X}_{N_b}) \psi_0(\mathbb{X}_{N_a}',\mathbb{X}_{N_b}) d\mathbb{X}_{N_b},
		\end{split}
	\end{equation}
	where $\mathbb{X}_{N_b} =\{\tilde{z}_1, \dots, \tilde{z}_N\}$ is the set of center-of-mass and relative coordinates on the set of particles of kind $b$. Now, if we perform the transformation on $\mathbb{X}_{N_a}$ to the center-of-mass and relative coordinates $z_1, \dots , z_N$, we arrive at
	\begin{equation}
		\begin{split}
			\rho_{N_a} \propto & \, e^{-\frac{\kappa_1}{2}\sum_{i=2}^N z_i^2}\widetilde{V}(z_2, \dots , z_N)\; e^{-\frac{\kappa_1}{2}\sum_{i=2}^N \tilde z_i^2}\widetilde{V}( z_2^\prime, \dots ,  z_N^\prime)\\
			&\times \int e^{-\frac{1}{4}\left(z_1+ \tilde z_1\right)^2} e^{-\frac{\kappa_2}{4}\left(z_1 - \tilde z_1 \right)^2}\; e^{-\frac{1}{4}\left( z_1^\prime+\tilde z_1 \right)^2} e^{-\frac{\kappa_2}{4}\left(z_1^\prime - \tilde z_1\right)^2} d\tilde z_1.
		\end{split}
	\end{equation}
	
	If we want to compute the purity of this reduced density matrix, the integral [see Eq. (\ref{rhopurity2})]
	\begin{equation}
		\text{Tr}\left( \rho _{N_a}^{2}\right) = \int \left[\rho_{N_a}(\mathbb{X}_{N_a},\mathbb{X}_{N_a}')\right]^2 \, d\mathbb{X}_{N_a} \, d\mathbb{X}_{N_a}'
	\end{equation}
	is easily performed in the coordinates $\{z_1, \dots , z_N\}$. We get 
	\begin{equation}
		\begin{split}
			\text{Tr}\left( \rho _{N_a}^{2}\right) =  \frac{2 \kappa_2}{\pi  (\kappa_2+1)}
			\int dz_1 d z_1^\prime \,\Big[& \int d\tilde z_1 e^{-\frac{1}{4}\left(z_1+\tilde z_1 \right)^2} e^{-\frac{\kappa_2}{4}\left(z_1 - \tilde z_1\right)^2}\\
			& \times e^{-\frac{1}{4}\left( z_1^\prime+\tilde z_1\right)^2} e^{-\frac{\kappa_2}{4}\left( z_1^\prime -\tilde z_1 \right)^2} \Big]^2,
		\end{split}
	\end{equation}
	which reduces to
	\begin{equation}
		\label{eq_purity_ab}
		\text{Tr}\left( \rho _{N_a}^{2}\right) =\frac{2\ \sqrt{\kappa _{2}}}{1+\kappa _{2}} =\frac{2 \left( 1-2N \Lambda \right)^{1/4}}{1+\sqrt{1-2N \Lambda}}. 
	\end{equation}
	
	\subsection{Single-particle reduced density matrix}
	
	We now present the procedure for determining the reduced density matrix $\rho_a$ of a single $a$-type particle (the case of a single $b$-type particle is entirely analogous). The matrix elements of $\rho_a$ are [following (\ref{int}) with $M_a=1,M_b=0$]
	\begin{equation}\label{rhosingle}
		\rho_a(x_1,x_1')= \int \psi_0(x_1,\mathbb X_{2N-1})\psi_0(x_1',\mathbb X_{2N-1}) \, d\mathbb X_{2N-1},
	\end{equation}
	with $\mathbb X_{2N-1}=\{x_2,\dots,x_N\}\cup \{x_{N+1},\dots,x_{2N}\}$. 
	In order to perform the integration we consider the center-of-mass and relative coordinates as in Eq. \eqref{eq_transformation_2}  for the sets of variables $\{x_2, \dots, x_N\}$ and $\{x_{N+1},\dots,x_{2N}\}$, and call the new variables $\{z_1, \dots , z_{N-1}\}$ and $\{\tilde z_1, \dots, \tilde z_N\}$, respectively.
	The ground state wave function thus takes the form
	\begin{flalign}\label{psi0a}
		\psi_0 = 
		& \, \mathcal{N}_N e^{-\frac{1}{4N}\left(x_1 + \sqrt{N-1}z_1 + \sqrt{N}\tilde z_1\right)^2} e^{-\frac{\kappa_2}{4N}\left(x_1 + \sqrt{N-1}z_1 - \sqrt{N}\tilde z_1\right)^2} \nonumber\\
		&\times e^{-\frac{\kappa_1}{2}\frac{N-1}{N}\left(x_1-\frac{z_1}{\sqrt{N-1}}\right)^2} e^{-\frac{\kappa_1}{2}\sum_{i=2}^{N-1}z_i^2} \, P(x_1, z_1,\dots, z_{N-1}) \nonumber\\
		&\times e^{-\frac{\kappa_1}{2}\sum_{i=2}^{N}\tilde z_i^{ 2}}\, \widetilde{V}(\tilde z_2, \dots, \tilde z_N),
	\end{flalign}
	where $P(u_1,\dots, u_N)$ denotes a polynomial. 
	
	The integration  in (\ref{rhosingle}) in terms of the new variables should be done in the following order: 
	first integrate with respect to $\{\tilde z_2, \dots, \tilde z_N\}$, and then with respect to $\tilde z_1$. Since this latter appears only in the exponential, the integral reduces to a Gaussian one. Then, to integrate over the variables $\{z_2,\dots, z_{N-1}\}$ we resort to
	\begin{equation}
		\label{eq:gauss-t-pot}
		\int_{-\infty}^{\infty} e^{-\kappa_1 z^2}z^{n}dz= 
		\begin{cases}
			0  &  n\text{ odd}, \\
			\kappa_1^{-\frac{n+1}{2}} \Gamma\left(\frac{n+1}{2}\right) & n\text{ even}.
		\end{cases}
	\end{equation}
	Finally, we will need to compute an integral of the form
	\begin{equation}
		\int_{-\infty}^{\infty} e^{-\alpha z^2 + \beta z} \sum_{k=0}^{2(N-1)}b_k z^k \,dz,
	\end{equation}
	with coefficients $\alpha$, $\beta$ and $b_k$'s depending on $x_1$, $x'_1$ and the interaction parameter. Using that
	\begin{equation}
		\begin{split}
			\mathcal{I}_{\alpha,\beta,n}:
			&=\int_{-\infty}^{\infty} e^{-\alpha z^2 + \beta z }z^{n}dz\\
			&=e^{\frac{\beta^2}{4 \alpha}} \sum_{k=0}^{\left[\frac{n}{2}\right]}\binom{n}{2k} \alpha^{-\left(k+\frac{1}{2}\right)}\left(\frac{\beta}{2 \alpha}\right)^{n-2k}\Gamma\left(k+\frac{1}{2}\right),
		\end{split}
	\end{equation}
	we are able to evaluate the reduced density matrix elements as
	\begin{equation}
		\label{rhosinglex}
		\rho_a(x_1,x'_1) = \sum_{k=0}^{2(N-1)}b_k (x_1,x'_1,\Lambda) \, \mathcal{I}_{\alpha,\beta,k}(x_1,x'_1,\Lambda).
	\end{equation}
	
	\section{Spatial correlations and entanglement}
	\label{sec_results}
	
	With the above results, we will now analyze the emergence of correlations in the hybrid system. 
	We start by focusing on the distribution functions of a single and a pair of (in)distinguishable particles, and then explore the entanglement correlations across different bipartitions of the system.
	
	\subsection{Single-particle and joint probability distributions}
	
	The study of single- and multi-particle correlations is of primary importance for the understanding of many-body systems, as they provide valuable information on the emergence of Bose-Einstein condensation, either of a particle or of a group of particles (in our case, a fermionic pair) \cite{yang_1962}. 
	Having outlined the procedure for calculating reduced density matrices, we will now examine the spatial correlations by considering the behaviour of their matrix elements in different interaction regimes.
	
	The single-particle reduced density matrix $\rho_a$ can be calculated straightforwardly for $N=1$ from the double Gaussian state of Eq.~\eqref{eq_solucion}
	\begin{equation}
		\psi_0(x_1,x_2)=\frac{\kappa_2^{1/4}}{\sqrt{\pi}}e^{-\frac{1}{4}(x_1+x_2)^2-\frac{\kappa_2}{4}(x_1-x_2)^2},
	\end{equation}
	leading to \cite{law_2005_cob} 
	\begin{equation}
		\label{eq:rhoa_N1}
		\rho_a(x,x')=\sqrt{\frac{2 \kappa_2}{\pi  (\kappa_2+1)}}e^{-\frac{\kappa_2}{2(1+\kappa_2)}(x+x')^2-\frac{1+\kappa_2}{8}(x-x')^2}.
	\end{equation}
	
	For higher number of pairs ($N\geq2$), we get for the matrix elements of $\rho_a$ the following expression, which is obtained with the method described in Sec. \ref{sect_rho_calculation} taking $M_a=1$ and $M_b=0$,
	\begin{equation}
		\rho _{a}( x,x') =e^{-\frac{1}{2}\left(\frac{x+x'}{2 \sigma _{+}}\right)^{2} -\frac{1}{2}\left(\frac{x-x'}{2 \sigma _{-}}\right)^{2}}\sum _{i=0}^{N-1}\sum_{j=0}^{2i} c_{ij}^{N} \ x^{2i-j} x^{\prime j},
		\label{eq:reddenmat-sigmas}
	\end{equation}
	with $c_{ij}^{N}=c_{ji}^{N}$ coefficients depending on $\Lambda$. Finding analytical expressions for these coefficients is a hard task, yet the widths $\sigma_+$ and $\sigma_-$ in the exponential terms can be straightforwardly found for arbitrary $N$ and $\Lambda$
	\begin{equation}
		\begin{split}
			&\sigma_+ = \left(\frac{4\kappa_1 \kappa_2 N}{\kappa_1+\kappa_2 [\kappa_1+2 (N-1)]}\right)^{-1/2},\\
			&\sigma_- = \left(\frac{\kappa_2+2 \kappa_1 (N-1)+1}{N}\right)^{-1/2} \,.\\
		\end{split}
		\label{eq_sigmas_rhoa}
	\end{equation}
	These quantities provide qualitative insight into the behavior of the diagonal and anti-diagonal elements of the reduced density matrix when varying the interactions strength $|\Lambda|$ for fixed $N$. They also bear information regarding the particle spatial distribution (diagonal) and the coherence terms (anti-diagonal). However, to better describe the diagonal and anti-diagonal behavior of the density matrix it is possible to define other widths which include a correction due to the polynomial term in (\ref{eq:reddenmat-sigmas}). In particular we calculate the standard deviation of the distributions along both directions, i.e. taking $x=x'$ and $x=-x'$. 
	Due to the symmetry of the reduced density matrix, the standard deviation $\sigma = \sqrt{\expval{x^2}-\expval{x}^2}$ reduces to $\sigma = \sqrt{\expval{x^2}}$, where $\expval{f(x)}$ denotes the integral of $f(x)$ weighted by a probability distribution. 
	
	As the diagonal part of $\rho_{a}$ represents a probability density, the quantity 
	\begin{equation}
		\sigma_a^{d} = \Big[\int x^2 \rho_a(x,x) dx\Big]^{1/2}
		\label{eq_exac_sig_a_d}
	\end{equation}
	is a natural choice for describing the spatial extension of this distribution. 
	To describe the spatial extension of the coherence, we notice that $\rho_a(x,-x)$ has $2(N-1)$ real roots ---as many as the degree of the polynomial part in \eqref{eq:reddenmat-sigmas}---, and $2N-1$ critical points corresponding to maxima or minima. The farthest of these critical points from the origin gives us an estimate of the spatial extension of the function $\rho_a(x,-x)$. We will thus estimate the width along the anti-diagonal direction of $\rho_a$ as 
	\begin{equation}
		\sigma_a^{ad}=\max \Big\{ \, x_c\in \mathbb{R} : \frac{d\rho_a(x,-x)}{dx}\Big|_{x_c}=0\Big\}.
		\label{eq_exac_sig_a_antid}
	\end{equation}
	
	Similarly, we can get an expression for the reduced density matrix of one pair of (distinguishable) particles $a$ and $b$, i.e. using the results of Sec. \ref{sect_rho_calculation} taking $M_a=1=M_b$ to obtain the following matrix elements of $\rho_{ab}$ 
	\begin{equation}
		\label{eq:rhoab}
		\begin{split}
			\rho_{ab}(x_a,x_b;x_a',x_b')=& \, e^{-\frac{1}{4}\left(\frac{x_{a} -x_{b} -x_{a} '+x_{b} '}{2\sigma _{1}}\right)^{2}} e^{-\frac{1}{4}\left(\frac{x_{a} +x_{b} -x_{a} '-x_{b} '}{2\sigma _{2}}\right)^{2}} \\
			&\times e^{-\frac{1}{4}\left(\frac{x_{a} -x_{b} +x_{a} '-x_{b} '}{2\sigma _{3}}\right)^{2}} e^{-\frac{1}{4}\left(\frac{x_{a} +x_{b} +x_{a} '+x_{b} '}{2\sigma _{4}}\right)^{2}}\\
			&\times \sum_{i=0}^{2(N-1)}\sum_{j,k,l=0}^{2i}c_{ijkl}^N \, x_a^{2i-j-k-l}x_b^jx_a'^kx_b'^{l} ,
		\end{split}
	\end{equation}
	where
	\begin{equation}
		\begin{split}
			&\sigma_1 = \left(\frac{2[\kappa_2+\kappa_1 (N-1)]}{N}\right)^{-1/2}  \text{ ,}\\
			&\sigma_2 = \left(\frac{2[\kappa_1 (N-1)+1]}{N}\right)^{-1/2} \text{ ,}\\
			&\sigma_3 = \left(\frac{2\kappa_1 \kappa_2 N}{\kappa_1+\kappa_2 (N-1)}\right)^{-1/2},\\
			&\sigma_4 = \left(\frac{2\kappa_1 N}{\kappa_1+N-1}\right)^{-1/2}.
		\end{split}
		\label{eq_sigmas_rhoab}
	\end{equation}
	
	The diagonal elements of $\rho_{ab}$ represent the joint probability distribution $\mathcal{D}_{ab}(x_a, x_b)=\rho_{ab}(x_a,x_b;x_a,x_b)$, which determines the probability of finding a fermion of species $a$ at position $x_a$ and a fermion of species $b$ at $x_b$. As stated below Eq. \eqref{eq_psi_in_ayb}, for vanishing interaction the state is separable, hence $\mathcal{D}_{ab}(x_a, x_b)$ factorizes as $\mathcal{D}_{ab}(x_a, x_b) = \rho(x_a) \rho(x_b)$ where $\rho(x)$ denotes the (marginal) spatial probability density for a single particle in a system of $N$ identical and non-interacting fermions \cite{lang2018correlations, deuretzbacher2007evolution} \footnote{When looking at the results of Refs. \cite{lang2018correlations, deuretzbacher2007evolution} it is important to keep in mind that in the cold atom and many body community it is usual to normalize the single-particle density matrix to the number of particles $N$ in order to obtain the density profile, here we adopt the use of the quantum information community and the normalization is made to 1.}. This means that in the non-interacting limit $\mathcal{D}_{ab}(x_a, x_b)$ has the same behavior in the direction $x_a=x_b$ as in $x_a=-x_b$. For finite interaction, the lack of this symmetry is an indicator of spatial correlations in $\rho_{ab}$, i.e, the more different is the density matrix across these two directions the more correlated is the system. We can invoke the spatial extension of $\mathcal{D}_{ab}$ along the rotated directions $(x_a+x_b)/\sqrt{2}$ and $(x_a-x_b)/\sqrt{2}$ to describe how $\mathcal{D}_{ab}$ behaves along the diagonal and anti-diagonal directions. Since $\mathcal{D}_{ab}$ is a two-dimensional probability distribution, we can compute these widths as
	\begin{equation}
		\begin{split}
			&\sigma_{ab}^d = \Big[\int \left(\frac{x_a+x_b}{\sqrt{2}}\right)^2 \mathcal{D}_{ab}\left(x_a,x_b\right) \,dx_a \,dx_b \Big]^{1/2}, \\
			&\sigma_{ab}^{ad} = \Big[ \int \left(\frac{x_a-x_b}{\sqrt{2}}\right)^2 \mathcal{D}_{ab}\left(x_a,x_b\right) \,dx_a \,dx_b\Big]^{1/2} \,.
		\end{split}
		\label{eq_exac_sig_ab}
	\end{equation}
	
	For the reduced density matrix of two particles of the same kind, $\rho_{aa}$ ($M_a=2$ and $M_b=0$ in Sec. \ref{sect_rho_calculation}), the calculations simplify resorting to the variables $R=(x_1+x_2)/\sqrt{2}$ and $r=(x_1-x_2)/\sqrt{2}$. Thus, we arrive at the following expression for the matrix elements
	\begin{equation}
		\label{eq:rhoaa}
		\begin{split}
			\rho _{aa} =& \, e^{-\frac{1}{2}\left(\frac{R+R'}{2\sigma _{p}}\right)^{2}} e^{-\frac{1}{2}\left(\frac{R-R'}{2\sigma _{m}}\right)^{2}} e^{-\frac{\kappa _{1}}{2}\left( r^{2} +r^{\prime 2}\right)}\\
			&\times r \, r^\prime\sum _{k=0}^{N-2}\sum _{l=0}^{k}\sum _{i=0}^{2( N-2) -k}\sum _{j=0}^{2i}\gamma^N_{i,j,k,l} R^{2i-j} R^{\prime j} r^{2( k-l)} r^{\prime 2l} \, .
		\end{split}
	\end{equation}
	with $\gamma_{i,j,k,l}^N$ being interaction-dependent coefficients, while
	\begin{equation}
		\begin{split}
			&\sigma_p = \left(\frac{2\kappa_1 \kappa_2 N}{\kappa_1+\kappa_2 (\kappa_1+N-2)}\right)^{-1/2} , \\
			&\sigma_m = \left(\frac{2[\kappa_2+\kappa_1 (N-2)+1]}{N}\right)^{-1/2} \, ,\\
		\end{split}
		\label{eq_sigmas_rhoaa}
	\end{equation}
	are the widths of the Gaussian dependence in the directions $x_1 + x_2 = x_1' + x_2'$, and $x_1 + x_2 = -(x_1' + x_2')$, respectively. The Gaussian parameter $\sigma_m$ characterizes the Gaussian decay of the coherences. It is also important to note the factors $(x_1 - x_2)$ and $(x_1' - x_2')$ in the polynomial part, which ensure the fulfillment of the Pauli exclusion principle since natural orbitals do not allow the simultaneous occupation of the same position by two fermions of the same species. If we take $x_1=x_1^\prime = x_a$ and $x_2=x_2^\prime = x_a^\prime$ we obtain the diagonal elements $\mathcal{D}_{aa}(x_a,x_a') = \bra{x_a,x_a'}\rho_{aa}\ket{x_a,x_a'}$, which represent the joint probability distribution of finding a particle of type $a$ in $x_a$, while another identical one is located at $x'_a$. To compare the latter probability with the diagonal spatial correlations $\mathcal{D}_{ab}$ of two particles of different kind, we can study the widths of $\mathcal{D}_{aa}(x_a,x_a^\prime)$ along the directions $(x_a+x_a')/\sqrt{2}$ and $(x_a-x_a')/\sqrt{2}$ by using the following widths:
	\begin{equation}
		\begin{split}
			&\sigma_{aa}^d = \Big[\int \left(\frac{x_a+x_a^\prime}{\sqrt{2}}\right)^2 \mathcal{D}_{aa}\left(x_a,x_a^\prime\right) \,dx_a \,dx_a^\prime \Big]^{1/2} ,\\
			&\sigma_{aa}^{ad} = \Big[ \int \left(\frac{x_a-x_a^\prime}{\sqrt{2}}\right)^2 \mathcal{D}_{aa}\left(x_a,x_a^\prime\right) \,dx_a \,dx_a^\prime \Big]^{1/2}.
		\end{split}
		\label{eq_exac_sig_aa}
	\end{equation}
	
	\subsubsection{Attractive regime}
	
	\begin{figure*}[tb]
		\centering
		\includegraphics[width=\columnwidth]{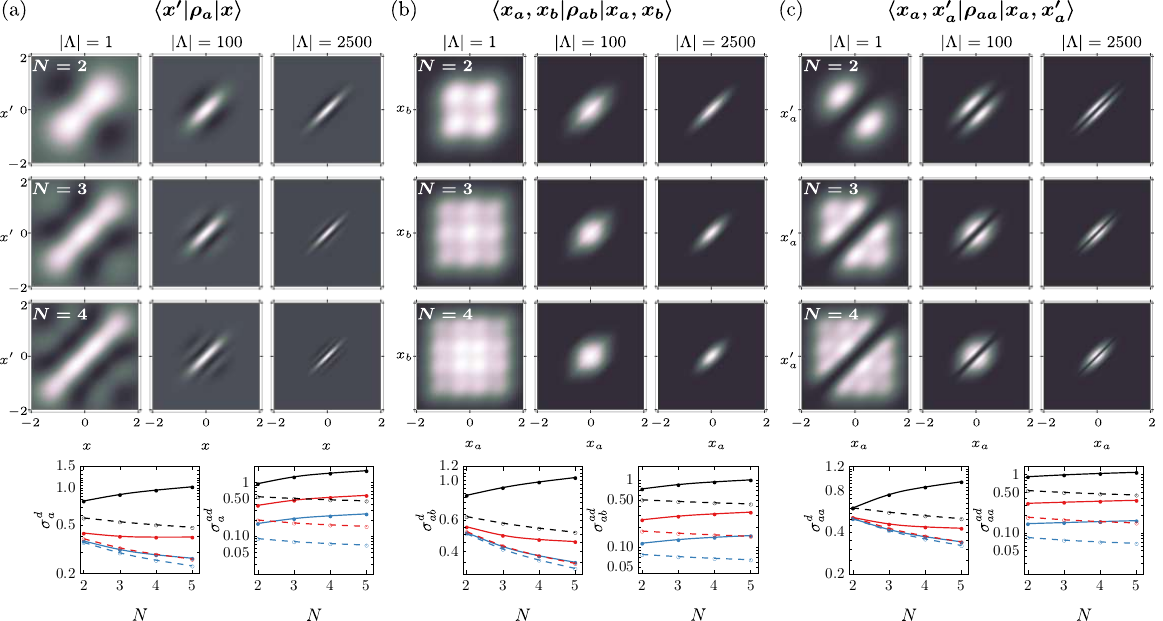}
		\caption{Matrix elements of the different computed density matrices in the attractive regime for $N=2,3,4$ (top to bottom) and $|\Lambda|=1, 100, 2500$ (left to right). (a) Single particle reduced density matrix, $\langle x'|\rho_a|x\rangle$, (b) Probability of finding a particle of kind $a$ in $x_a$ and a particle of kind $b$ in $x_b$, $\mathcal D_{ab}(x_a,x_b)=\langle x_a,x_b|\rho_{ab}|x_a,x_b\rangle$, and (c) Probability of finding a particle of kind $a$ in $x_a$ and another particle of the same type in $x'_a$, $\mathcal D_{aa}(x_a,x'_a)=\langle x_a,x'_a|\rho_{aa}|x_a,x'_a\rangle$. The bottom row depicts the diagonal ($\sigma^{d}_{\circ}$) and anti-diagonal ($\sigma^{ad}_{\circ}$) widths as a function of $N$ in log scale for $|\Lambda| = 1,\, 100,\, 2500$ (black, red, and blue curves respectively). Two different measures for the widths are presented, the ones associated to the Gaussian decays given in Eqs.~\eqref{eq_sigmas_rhoa}, \eqref{eq_sigmas_rhoab} and \eqref{eq_sigmas_rhoaa} (empty dots with dashed lines) and the corrected widths of Eqs. \eqref{eq_exac_sig_a_d}, \eqref{eq_exac_sig_a_antid}, \eqref{eq_exac_sig_ab}, and \eqref{eq_exac_sig_aa} (dots with solid lines).} 
		\label{fig_rhos_at}
	\end{figure*}

	Figure \ref{fig_rhos_at} (a) shows the matrix elements $\bra{x} \rho_a \ket{x'}$ in the attractive regime for different values of  $N$ (rows from top to bottom corresponding to $N=2,3,4$) and  $\Lambda$ (columns from left to right corresponding to $|\Lambda|= 1, \,100, \,2500$). The color scale is such that the value increases as the color becomes brighter. For weak interactions, the results are in agreement with those obtained for a gas of identical non-interacting fermions \cite{girardeau2001ground}. As expected, for a fixed $N$ as the strength of the attractive interaction increases the distribution becomes more localized along the diagonal direction $x'=x$ indicating an increase in the spatial correlations. When increasing the number of pairs $N$, we observe that for small $|\Lambda|$ the distribution extends with $N$ while for larger $|\Lambda|$ it tends to concentrate towards the origin.

	The bottom row of Figure \ref{fig_rhos_at} (a) depicts two different width measures as a function of the number of pairs and for $|\Lambda|=1$ (black), $|\Lambda|=100$ (red), and $|\Lambda|=2500$ (blue). One measure is given in terms of the Gaussian widths in Eq.~\eqref{eq_sigmas_rhoa} (non-filled dots and dashed guiding lines), and the second one are the quantities given in Eqs.~\eqref{eq_exac_sig_a_d} and \eqref{eq_exac_sig_a_antid} (dots and solid lines). This second width measure corrects the first one by including the fermionic exchange effects given by the polynomial term in the density matrix expansion and are in agreement with the behavior depicted in the upper color panels: for $|\Lambda| = 1$ the width increases with $N$ while for $|\Lambda|=100$ and $2500$ it decreases when increasing the number of pairs in the depicted range. Since the Gaussian widths along the diagonal and the anti-diagonal directions decrease when increasing both $N$ and $\Lambda$ (with a faster decrease in the diagonal direction) we conclude that the Gaussian widths provide a qualitative description for the system's behavior when increasing $|\Lambda|$ in the attractive regime at fixed $N$ but not about its behavior when increasing $N$ for a fixed interaction. This means that for fixed number of pairs the Gaussian part of the density matrix defined by the harmonic confinement dominates over the polynomial part but when increasing $N$ the fermionic exchange effects synthesized in the polynomial term cannot be neglected.
	
	Figure \ref{fig_rhos_at} (b) illustrates the distribution $\mathcal{D}_{ab}(x_a, x_b)$ for different values of the interaction parameter and number of particle pairs ($N=2,3,4$ in rows from top to bottom and $|\Lambda|=1,\,100,\,2500$ in columns from left to right). If the state is separable the joint particle distribution factorizes and due to the symmetry of the external trap the distribution should have a similar profile along the diagonal and anti-diagonal directions, as can be observed in the left column of Fig. \ref{fig_rhos_at} (b) corresponding to a weak interaction strength for which the profiles show oscillations with $N^2$ peaks in agreement with known results for identical non-interacting fermions \cite{deuretzbacher2007evolution}. The highest probability (brighter points) is associated to the nearest neighbor position in an equidistant array, which for even $N$ does not include the central position whereas for odd $N$ it does. For $N=2$ and $4$ the most likely positions are $x_a = \pm x_b$, while for $N=3$ it is $x_a=x_b=0$. As expected for weak interaction strengths, the similarity between the diagonal and the anti-diagonal directions of $\mathcal D_{ab}$ reflect low correlations between particles, on the contrary, as the interaction strength increases for fixed $N$ the distribution reaches its maximum (minimum) value along the diagonal (anti-diagonal) direction within an effective length that decreases as the interaction strength increases. This suggests that fermions of different species coupled via a harmonic interaction tend to be closer as the attraction increases, allowing them to occupy nearby positions that accumulate towards the origin for sufficiently large $|\Lambda|$. By comparing the rows in Fig. \ref{fig_rhos_at} (b), we note that an increase in the number of particles for fixed $\Lambda$ weakens the spatial correlations as evidenced by more symmetric distributions along the diagonal and anti-diagonal directions. The bottom row of figure \ref{fig_rhos_at} (b) shows two measures of the diagonal and anti-diagonal widths of the joint distribution as a function of $N$ and for different values of the interaction strength ($|\Lambda|=1$ in black, $|\Lambda|=100$ in red, and $|\Lambda|=2500$ in blue). The Gaussian widths $\sigma_3$ and $\sigma_4$ of Eq.~\eqref{eq_sigmas_rhoab} are shown as empty dots with dashed lines while the standard deviations of Eq.~\eqref{eq_exac_sig_ab} are depicted as dots with solid lines. As in the case of the single particle distributions, the Gaussian widths provide for a good qualitative understanding of the behavior of $\mathcal{D}_{ab}(x_a, x_b)$ for a fixed number of pairs $N$ but fail to describe the behavior of the system when increasing $N$ for a fixed interaction strength.

	Figure \ref{fig_rhos_at} (c) shows the diagonal matrix elements $\langle x_a,x'_a|\rho_{aa}|x_a,x'_a\rangle$ representing the joint probability distribution $\mathcal{D}_{aa}(x_a,x'_a)$ which gives the probability of finding a particle of type $a$ in $x_a$ while another identical particle is located at $x'_a$. As expected due to the Pauli exclusion principle the correlations vanish at the contact point $x_a=x_a^\prime$. For weak interactions, the results are in agreement with those known for a gas of identical non-interacting fermions \cite{girardeau2001ground}. As the interaction strength increases, the extent of the distribution along the diagonal direction remains of the order of the harmonic trap being determined by the distribution of the center of mass of the system, while the extent along the anti-diagonal direction decreases. 
	The root-mean-square separation between two identical fermions (given by the square root of the expectation value of $(x - x^\prime)^2$ for which it is necessary to take $x_1=x_1^\prime=x$ and $x_2=x_2^\prime=x^\prime$ in Eq. \eqref{eq:rhoaa} and equate the primed variables to the unprimed ones) gives $(2\kappa_{1})^{-1/2} = [4(1 - N\Lambda)]^{-1/4}$, showing that as the strength of the attractive interaction increases fermions of the same species arrange at small distances compared to the extent of the external trap without ever coinciding in the same position due to Pauli repulsion. In other words, under a strongly attractive harmonic interaction all identical particles in the system remain spatially bound with a microscopic separation induced by Pauli repulsion depicting a very different behavior compared to the one observed for fermions under a strongly attractive contact interaction that form strongly bound pairs which remain separated by a distance comparable to the amplitude of the harmonic trap \cite{Jimenez2023}.

	As before, the bottom row of Fig. \ref{fig_rhos_at} (c) depicts two measures of the diagonal and anti-diagonal widths of the joint distribution as a function of the number of pairs $N$ and for different interaction strengths $|\Lambda|$. The Gaussian width $\sigma_p$ of Eq.~\eqref{eq_sigmas_rhoaa} and the quantity $(2\kappa_{1})^{-1/2} = [4(1 - N\Lambda)]^{-1/4}$ are depicted as non-filled dots with dashed lines while the standard deviations of Eq.~\eqref{eq_exac_sig_aa} are shown as full dots and solid lines. As in the case of the single particle reduced density matrix and the joint probability distribution for particles of different kinds, the Gaussian widths qualitatively capture the behavior of the system when varying the interaction strength for a fixed number of pairs $N$.
	
	Our results suggest that particles of the same species have a larger characteristic separation than those of different species, as can be seen when comparing the extension along the anti-diagonal direction of the distributions $\mathcal{D}_{aa}$ and $\mathcal{D}_{ab}$ for equal values of $|\Lambda|$ and $N$. Therefore, the most likely arrangement when measuring the positions of all particles is an alternating (or antiferromagnetic, if we interpret fermionic species as spin states) configuration inside the trap. To conclude the analysis of the spatial correlations in the attractive regime, we confirm the absence of off-diagonal long-range order (ODLRO) \cite{yang_1962, astrakharchik_2005_prl95, BEC_stringari}. Following the approach in \cite{yang_1962}, a reduced two-particle density matrix exhibits ODLRO in the coordinate representation whenever the two-particle reduced density matrix remains non-zero for unprimed coordinates microscopically close to a localization $x$, and primed coordinates close to another localization $x^\prime$, with $x$ and $x^\prime$ macroscopically separated. By setting $x_a = x_b = x $ and $ x_a^\prime = x_b^\prime= x^\prime $ in Eq.~\eqref{eq:rhoab} we obtain a Gaussian dependence $ e^{-(x - x')^2/\sigma_{2}^2}$ imposed by the harmonic confinement and which precludes the emergence of ODLRO despite the fact that as the attraction increases all particles tend to cluster within a single localized region preserving Pauli repulsion between fermions of the same species.
	
	\subsubsection{Repulsive regime}
	
\begin{figure*}[tb]
		\centering
		\includegraphics[width=\textwidth]{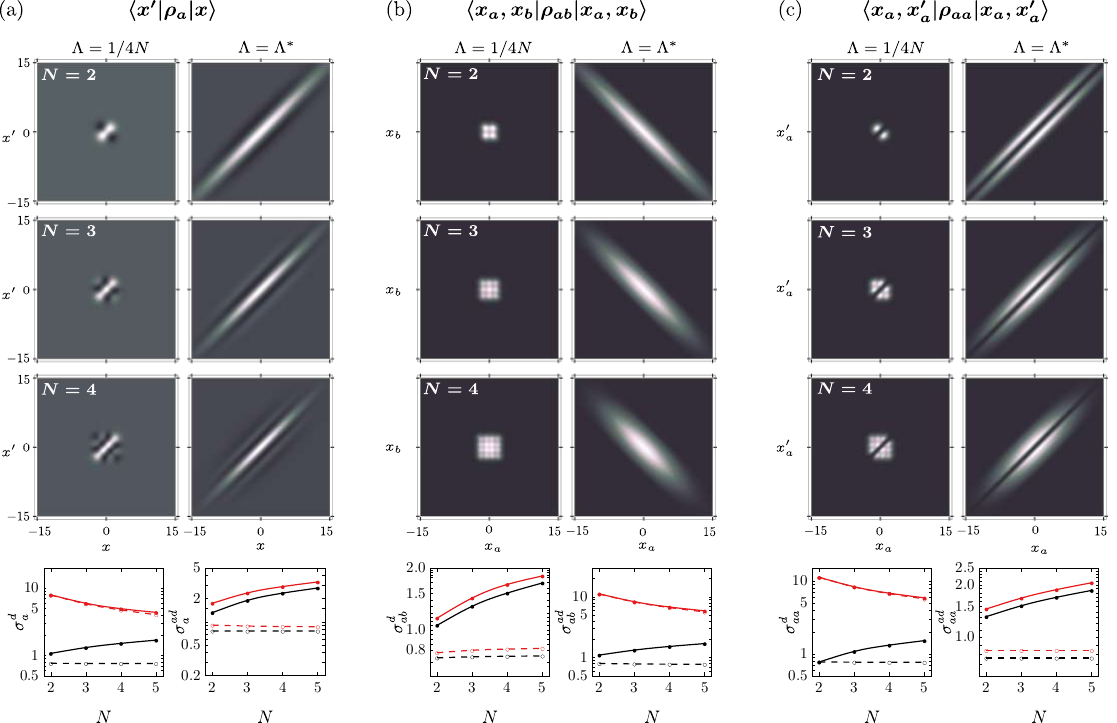}    
		\caption{Matrix elements of the calculated density matrices for $N=2,3,4$ (top to bottom) and repulsive interactions $\Lambda =1/(4N)$ and $\Lambda =\Lambda^*=1/(2N)-1/10^6$ (left to right). (a) Single particle reduced density matrix, $\langle x'|\rho_a|x\rangle$, (b) Joint probability distribution $\mathcal D_{ab}(x_a,x_b)=\langle x_a,x_b|\rho_{ab}|x_a,x_b\rangle$, and (c) Joint probability distribution $\mathcal D_{aa}(x_a,x'_a)=\langle x_a,x'_a|\rho_{aa}|x_a,x'_a\rangle$. The bottom row depicts the diagonal ($\sigma^{d}_{\circ}$) and anti-diagonal ($\sigma^{ad}_{\circ}$) widths as a function of $N$. The Gaussian widths given by Eqs.~\eqref{eq_sigmas_rhoa}, \eqref{eq_sigmas_rhoab} and \eqref{eq_sigmas_rhoaa} (empty dots and dashed lines) together with the standard deviations given in Eqs. \eqref{eq_exac_sig_a_d}, \eqref{eq_exac_sig_a_antid}, \eqref{eq_exac_sig_ab}, and \eqref{eq_exac_sig_aa} (dots with solid lines as guide to the eye). The black and red curves correspond respectively to $\Lambda =1/(4N)$ as a week repulsive interaction strength and to $\Lambda =\Lambda^*=1/(2N)-1/10^6$ which is close to the limit where the system does not support a bound state anymore.}
		\label{fig_panel_rep}
	\end{figure*}

	
	We analyze now the behavior of the distributions in the repulsive regime. Figure \ref{fig_panel_rep} shows the matrix elements of $\rho_a$, $\rho_{ab}$, and $\rho_{aa}$ for $N=2,3,$ and 4 (rows), and $\Lambda =1/(4N)$ and $\Lambda =\Lambda^*=1/(2N)-1/10^6$ (columns). The two measures for the width of the distributions along the diagonal and anti-diagonal direction are shown in the bottom row of Fig. \ref{fig_panel_rep}. As before the Gaussian widths of Eqs.~\eqref{eq_sigmas_rhoa}, \eqref{eq_sigmas_rhoab} and \eqref{eq_sigmas_rhoaa} are depicted as empty dots with dashed lines while the standard deviations given in Eqs. \eqref{eq_exac_sig_a_d}, \eqref{eq_exac_sig_a_antid}, \eqref{eq_exac_sig_ab}, and \eqref{eq_exac_sig_aa} are shown as dots with solid lines as guide to the eye, also, black curves correspond to $\Lambda =1/(4N)$ while the red color is used for $\Lambda =\Lambda^*=1/(2N)-1/10^6$. In most of the repulsive interval $0<\Lambda\leq 1/(2N)$ the spatial correlations do not differ significantly from those obtained with weak interactions, however, as $ N\Lambda$ approaches the limit  $1/2^-$ the correlations undergo an abrupt change. Due to the competition between particle repulsion and the confinement potential the probability distributions depicted in Fig. \ref{fig_panel_rep} have a spatial extension several times larger than the one of the external trap. Also, the width of the distributions increase with the interaction strength and diverge in the limit $N\Lambda \to 1/2^-$ (as can be seen from Eq.~\eqref{eq_solucion} the characteristic separation between the centers of mass of the two subsystems having particles of the same species varies as $\kappa_2^{-1/2}=(1-2N\Lambda)^{-1/4}$) where the system is no longer capable of sustaining bound states.
	
	The single particle reduced density matrix $\rho_a(x, x')$, shown in Fig. \ref{fig_panel_rep} (a), exhibits a pronounced localization around the diagonal and minor oscillations in the anti-diagonal direction associated to the coherences. For increasing interaction strength this localization around the diagonal direction increases with an increasing associated width. For weak repulsive interaction the probability to find a particle of kind $a$ in $x_a$ and a particle of kind $b$ in $x_b$ depicts a similar behavior than the one found for weak attractive interaction, however, for large interaction strength repulsive particles are more likely to be found in the anti-diagonal direction $x_a = -x_b$ with the highest probability concentrated around the origin. We interpret this as a consequence of the trade off between the strong repulsion and confinement, pushing the system into an energy preferred configuration having the fermions of different species closed together with a spacial width that decreases when the number of pairs increases. The probability distribution for particles of the same species follows a similar behavior, with the difference being that for large interaction strength the probability spreads parallel to the diagonal direction following the Pauli exclusion principle and with a characteristic width that increases with $N$, see Fig. \ref{fig_panel_rep} (c). For the depicted $N$ and $\Lambda$ values in the repulsive regime we found that both particles of the same and different species maintain a characteristic separation of approximately the size of the harmonic trap, also, the Gaussian widths obtained along the direction with the highest spreading degree (diagonal for $\rho_a$ and $\rho_{aa}$, and anti-diagonal for $\rho_{ab}$) are in good agreement with the corrected ones in the limit of strong interaction.
	
	ODLRO is also absent in this regime. When increasing the interaction strenght the system undergoes a spatial expansion, with particles of the same species forming separate domains. However, even when the system's spatial extent grows significantly for $N\Lambda \approx 1/2$, the coherences of the single- and two-body density matrices remain finite and on the order of the length scale of the external trap. The persistence of finite coherence lengths confirms the absence of ODLRO in the strongly repulsive regime and can be explicitly verified by taking the limits \( \kappa_2 \to 0 \) and \( \kappa_1 \to \sqrt{1/2} \) in Eqs. \eqref{eq_sigmas_rhoa}, \eqref{eq_sigmas_rhoaa}, and \eqref{eq_sigmas_rhoab}.   
	
	\subsection{Entanglement for different bipartitions}
	
	By computing the purity of the reduced density matrix according to the exposition in Section \ref{sect_rho_calculation}, we are able to analyze the entanglement across different bipartitions in the hybrid system resorting to the entanglement measure given in Eq.~\eqref{eq_e_bip_mix_Npairs}. In the light of Eq.~ \eqref{eq_energyMIN}, the parameter $N \Lambda$ is a natural one for studying the system, therefore, we will analyze the amount of entanglement (excluding the anti-symmetry correlations as explained in Sec. \ref{sec_entanglement}) as $N |\Lambda|$ varies for the attractive and repulsive case.

	We will focus on the entanglement between one or two particles (of the same or different species) and the rest of the system. 
	First, the entanglement between a single particle of kind $a$ and the rest of the system is denoted by $\varepsilon_a=\varepsilon_{1|(2N-1)}$ (that is, $\varepsilon_{(M_a+M_b)|(N_a-M_a+N_b-M_b)}$ in Eq. (\ref{eq_e_bip_mix_Npairs}) with $M_a=1$, $M_b=0$, $N_a=N_b=N$) and is calculated as 
	\beq
	\varepsilon_a = 1 - N \, \textrm{Tr}(\hat \rho^2_{a}).
	\eeq
	Notice that due to the symmetry of the problem, $\varepsilon_a$ equals $\varepsilon_b$, so the nature of the isolated particle is irrelevant. Second, the entanglement across a bipartition that isolates a pair of particles of different species from the remaining $2N-2$ parties, denoted as $\varepsilon_{ab}$, corresponds to $\varepsilon_{(M_a+M_b)|(N_a-M_a+N_b-M_b)}$ in Eq. (\ref{eq_e_bip_mix_Npairs}) with $M_a=M_b=1$, $N_a=N_b=N$, and is given by 
	\beq
	\varepsilon_{ab}=1-N^2 \, \textrm{Tr}(\hat \rho^2_{ab}).
	\eeq
	Third, the entanglement between two identical particles and the remaining $2N-2$ parties, corresponding to  $\varepsilon_{(M_a+M_b)|(N_a-M_a+N_b-M_b)}$ in Eq. (\ref{eq_e_bip_mix_Npairs}) with $M_a=2$, $M_b=0$, and $N_a=N_b=N$, will be denoted as $\varepsilon_{aa}$, and is given by 
	\beq
	\varepsilon_{aa} = 1 - \frac{N(N-1)}{2} \textrm{Tr}(\hat \rho^2_{aa}).
	\eeq
	Finally, we consider the entanglement in the bipartition that divides the system into particles of different types, corresponding to Eq.~\eqref{eq_e_bip_mix_Npairs} with $M_a=N_a=N$ and $M_b=0$ and denoted as $\varepsilon_{N_a|N_b}$ (stressing with this notation that the $N$ parties in each subsystem are of the same species). From Eq.~\eqref{eq_purity_ab} and the discussion below Eq.~\eqref{eq_e_bip_mix_Npairs} we get 
	\begin{equation}
		\varepsilon_{N_a|N_b}=1-\textrm{Tr}(\rho^2_{N_a})
		=\frac{1+\sqrt{1-2N\Lambda}-2(1-2N\Lambda)^{1/4}}{1+\sqrt{1-2N\Lambda}} \,.
	\end{equation}
	
	\begin{figure*}[tb]
		\centering
		\includegraphics[width=\textwidth]{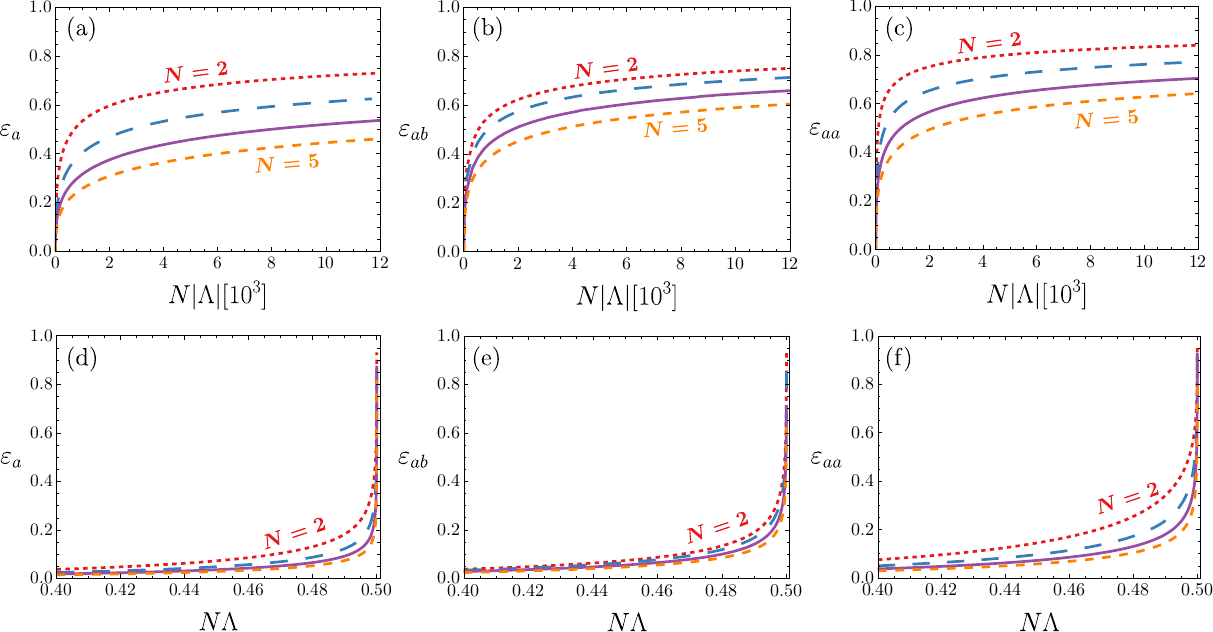}
		\caption{Entanglement between one particle and the rest of the system $\varepsilon_{a}$, between one pair of different species and the remaining $2N-2$ particles $\varepsilon_{ab}$, and entanglement between two particles of kind $a$ and the remaining $2N-2$ particles $\varepsilon_{aa}$ as a function of $N |\Lambda|$ for $N=2,\,3,\,4\,,5$ (red dotted, blue large-dashed, purple solid, and orange medium-dashed lines --from top to bottom). The attractive regime is depicted in panels (a)-(c) while the entanglement measures obtained for repulsive interactions are shown in panels (d)-(f).}
		\label{fig_ent}
	\end{figure*}
	
	Figure \ref{fig_ent} depicts $\varepsilon_a$, $\varepsilon_{ab}$ and $\varepsilon_{aa}$ for the attractive [panels (a)-(c)] and repulsive [panels (d)-(f)] cases. All the entanglement measures are shown as a function of the rescaled parameter $N |\Lambda|$ and for $N=2,\,3,\,4\,,5$ (red dotted, blue large-dashed, purple solid, and orange medium-dashed lines --from top to bottom). For both attractive and repulsive interaction, the entanglement measures are monotonically increasing functions of the interaction strength. This phenomenon has already been observed in other confined systems with harmonic interaction, such as in the $N$-harmonium \cite{benavides2014entanglement} and in a mixture of bosonic species \cite{alon2024properties}. Also, in all cases smaller systems (lower values of $N$ for a fixed interaction strength) favor a larger amount of entanglement. This can be understood as follows: for both the repulsive and attractive interactions, as $N$ increases the energy tends to that of a binary non-interacting (hence separable) system, with vanishing entanglement (see Eq.~\eqref{eq_energyMIN} and discussion therein). It is therefore reasonable to expect that any entanglement measure (except when considering entanglement among the species $a$ and $b$) decreases as $N$ increases, ultimately vanishing for sufficiently large $N$, in line with the behavior captured in Fig. \ref{fig_ent}. This, in turn, is consistent with the results in Ref. \cite{alon2024properties} for a bosonic system, where the authors observed that the mean field solution (that erase the correlations)  and the exact one give the same energy in the thermodynamic limit.
	
	In the attractive regime the entanglement rapidly increases in the vicinity of $\Lambda =0$ (non-interacting system) and increases more slowly for large $|\Lambda|$, whereas in the repulsive case an opposite behavior is observed: the entanglement enhances slowly near the non-interacting regime and extremely fast as $N \Lambda \to 1/2^-$, reaching its maximum value ($\varepsilon=1$) precisely at $N\Lambda=1/2$. 
	
	\begin{figure}[htb]
		\centering
		\includegraphics[width=0.95\textwidth]{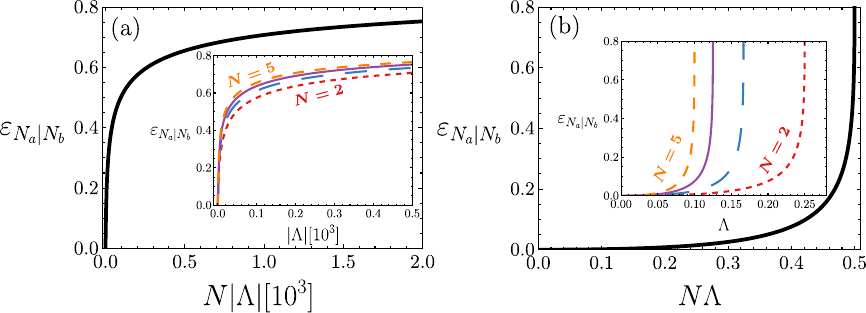}
		\caption{Entanglement between all particles of type $a$ and all particles of kind $b$ (a) in the attractive regime ($\Lambda>0$) and (b) for repulsive interactions, i.e. $0<\Lambda \leq 1/(2N)$. All the curves for different values of $N$ collapse when plotted against the rescaled interaction parameter $|\Lambda| N$, the inset depict the curves and for $N=2,\,3,\,4\,,5$ (red dotted, blue large-dashed, purple solid, and orange medium-dashed lines, respectively --from outermost right to left) plotted against the interaction parameter $|\Lambda|$.}
		\label{fig_ent_ABall}
	\end{figure}
	
	The entanglement between all the particles of one species and all the particles of the other species is shown in Fig.~\ref{fig_ent_ABall} for the attractive (a) and repulsive (b) regimes. The curves of the entanglement as a function of $|\Lambda|$ for different $N$ ($N=2,\,3,\,4\,,5$ in the inset, same color code as before) collapse into a single curve exhibiting a monotonically increase of the entanglement with respect to the rescaled interaction parameter. For the non-interacting case ($\Lambda = 0$) the entanglement is zero, as expected. 
	
	In the attractive regime and for $N|\Lambda| \gg 1$, the entanglement reduces to $\varepsilon_{N_a|N_b}=1-2/(2N |\Lambda|)^{1/4}$, so in the strong attractive limit, or in the attractive regime in systems involving a large number of particles, the entanglement between the composites of different species becomes practically maximal. In the repulsive regime the entanglement increases extremely rapidly, so as $\Lambda$ approaches $1/(2N)$, the entanglement becomes $\varepsilon_{N_a|N_b}=1-2(1-2N\Lambda)^{1/4} \approx 1$. 
	
	The insets of Fig.~\ref{fig_ent_ABall}  show that the entanglement between subsets of different species increases as $N$ increases, for fixed interaction strength. This behavior is the opposite to the trend observed for $\varepsilon_a$, $\varepsilon_{ab}$, and $\varepsilon_{aa}$, and can be understood based on the screening effect suggested by the behavior of the energy for large $N$. Taking into account that the interactions occur only between particles of different kind ($ab$), the effective interaction between particles of the same species appears as a second-order effect mediated by a third different particle [$a(b)a$] or [$b(a)b$]. We conjecture that the mediated interaction between particles of the same kind is strongly screened when the number of particles increases while the interaction $ab$ is less affected. Since the correlations between particles of the same kind are either exchange correlations (which are substracted in our entanglement definition) or a consequence of this second order mediated interaction which is being screened when $N$ increases, this would explain both the decrease of $\varepsilon_a$, $\varepsilon_{ab}$ and $\varepsilon_{aa}$ (corresponding to bipartitions that include particles of the same kind in a given subsystem) and the increasing behavior of $\varepsilon_{N_a|N_b}$ (corresponding to the only bipartition that excludes screening second order interactions for the subsystems) when $N$ increases.
	
	
	As a final comment, from the ground state energy expression [see Eq.~\eqref{eq_energyMIN}] we also estimate the fraction of the available volume which is used by the system. Our reasoning pointed that in the non-interacting case the system effectively takes all the available volume, in the strongly repulsive limit the system natural scale competes with the one of the trap, and in the strongly attractive regime (or attractive with large $N$) the volume occupied by the system vanishes as $1/\sqrt{ N |\Lambda|}$. Based on some results of Refs. \cite{baccetti_2013,chudzicki_2010,law_2005_cob} the authors of Ref. \cite{cuestas_2020} have related the availability of physical space with the availability of states in the state space and the presence of entanglement. Our present analysis supports this hypothesis in an exactly solved system: in the attractive case we confirm that when the confinement volume is much larger than the effective volume of the system, the different studied bipartitions are strongly entangled. Moreover, in the non-interacting case the system occupies all the available volume and the entanglement vanishes. However, in the repulsive regime the entanglement increases even when the scale of the system is progressively loosing available space. This could mean either that the availability of volume is a necessary but not sufficient condition for the presence of entanglement in confined systems, or that strong entanglement arises whenever one of the two scales is much larger than the other one, and not necessarily when the volume of the trap is much larger than the effective volume of the system. 
	
	\section{Summary and Conclusion}
	\label{sec_conclusions}
	
	Using an exactly solvable model, we advanced in the study of spatial distributions and correlations in a binary composite of fermionic species confined in a one-dimensional harmonic trap, providing valuable insights into complex many-body phenomena. In particular, we considered the balanced composite of $2N$ fermions, $N$ of which are (indistinguishable) parties of the species $a$ while the remaining $N$ parties are (indistinguishable) elements of the second species $b$, subject to a harmonic Hamiltonian with an interaction term that couples particles of different kind. The exact solution for the ground state, appropriately symmetrized in order to fulfill the antisymmetrization requirement under the exchange of any two particles of the same species, is obtained.  
	
	A significant contribution of our work is the introduction of an entanglement measure specifically designed for hybrid systems comprising both distinguishable and identical fermions, of arbitrary number. Our measure reduces to the standard entropy-based entanglement measure when the bipartition divides the system into distinguishable subsystems ($a$-type and $b$-type particles), whereas for any other bipartition it is suitably adapted to exclude exchange correlations among parties of the same kind. In this way, the proposed measure quantifies only useful entanglement (beyond Slater correlations), thus enabling precise quantification of entanglement in quantum information applications that involve hybrid composites. Besides exploring the entanglement correlations across different bipartitions of the system, we also analyze the distribution functions of a single and a pair of (in)distinguishable particles delving into their spatial distribution and correlations, in different interacting regimes and for varying number of particle pairs $N$. 
	
	The analysis of the density matrix elements revealed detailed information about the spatial extent of the system, the characteristic separation between particles of the same and different species, the coherence properties of the single-particle reduced state, and the spatial correlations. In the strongly attractive regime, all particles are spatially confined within a small localized region, while preserving the Pauli repulsion between
	fermions of the same species, favoring an alternating configuration between particles of different species. 
	In the strong repulsive regime, i.e. when the ratio between the interaction and confining strength is close to the particle density $\Lambda \approx 1/(2N)$, the interaction favors the formation of two spatially separated regions, one containing particles of type $a$ and the other one containing particles of type $b$. The extent of these domains is comparable to the width of the external trap, and the characteristic separation between them exhibits an abrupt increase as the interaction strength approaches the limit $\Lambda =1/(2N)$, beyond which the system does not support a bounded state. 
	
	The entanglement across the bipartition that divides fermions of different species is fully determined by the extensive parameter $ N\Lambda$ and exhibits an increasing behavior both in this parameters as in $N$ (for a fixed interaction strength). In contrast, the entanglement across bipartitions in which one of the subsystems contains indistinguishable parties depends not only on $ N\Lambda$ but also on $N$, and displays a decreasing tendency as $N$ increases. We conjecture that this latter feature ensues from a screen effect that weakens the effective interaction between identical particles (mediated by a third, distinguishable, one) in either side of the bipartition. Such behavior provides insights into how entanglement is distributed across the system under different partitioning schemes. Overall, our findings demonstrate, as expected, that the degree of entanglement increases with interaction strength in both attractive and repulsive regimes. 
	
	Our results deepen our understanding of the intricate interplay between interactions, indistinguishability, and entanglement in multi-particle quantum systems.

	\ack{
		M.D.J. and A.P.M. acknowledge funding from Grants No. PICT 2020-SERIEA-00959 from ANPCyT (Argentina) and No. PIP 11220210100963CO from CONICET (Argentina) and partial support from SeCyT, Universidad Nacional de Córdoba (UNC), Argentina. A.V.H. acknowledges financial support from DGAPA-UNAM through project PAPIIT IN112723. E.C. was supported by JSPS KAKENHI grant number JP23K13035 and the Horizon Europe programme HORIZON-CL4-2022-QUANTUM-02-SGA via the project 
		\href{https://doi.org/10.3030/101113690}{101113690} (PASQuanS2.1). 
	}
	
	\appendix 
	
	\section{Ground state's normalization constant}
	\label{sec_app_normalization}
	
	In the present appendix we present the calculation of $\mathcal N_N$, the normalization constant for the ground state of $N$ pairs. If we perform the integration of $\displaystyle |\psi_0 |^{2}$ over the spatial coordinates, we should impose the condition 
	\begin{equation*}
		1=\int |\psi_0 (x_{1}, \dots, x_{2N})|^{2} dx_{1} \cdots dx_{2N}.
	\end{equation*}
	By using Eq. \ref{eq_solucion}, and replacing $x_i$ by $\kappa_1^{-1/2}y_i$, we get
	\begin{equation}
		\begin{split}
			1 = & \, \mathcal{N}_{N}^{2} \ \kappa _{1}^{-N} \kappa _{1}^{-N( N-1)}\int e^{-\frac{1}{2N\kappa _{1}}\left(\sum\limits_{i=1}^{2N} y_{i}\right)^{2}} \\
			&\times e^{-\frac{\kappa _{2}}{2N\kappa _{1}}\left(\sum\limits_{i=1}^{N} y_{i} -\sum _{j=N+1}^{2N} y_{j}\right)^{2}}\\
			&\times  \prod _{1\leq i< j\leq N} e^{-\frac{1}{N}( y_{i} -y_{j})^{2}}( y_{i} -y_{j})^2\\
			&\times \ \prod _{N+1\leq i< j\leq 2N} e^{-\frac{1}{N}( y_{i} -y_{j})^{2}}( y_{i} -y_{j})^2 \ dy_{1} \cdots dy_{2N}.
		\end{split}
	\end{equation}
	Then, let us consider the variables
	\begin{equation}
		\begin{split}
			&Y_1 = \frac{\kappa_1^{-1/2}}{\sqrt{2N}}\sum_{j=1}^{2N} y_j,\\
			&Y_{2N} = \left(\frac{\kappa_2}{\kappa_1}\right)^{1/2} \frac{1}{\sqrt{2N}}\left(\sum_{j=1}^{N} y_j - \sum_{j=N+1}^{2N} y_j\right),\\
		\end{split}
	\end{equation}
	and $Y_i = \sum_{j=1}^{2N} v^{(i)}_jy_j$, with $\{v^{(i)}\}$ defined as in Eq. (\ref{eq:basis_elements}), for $i=2,\dots,2N-1$. By taking into account Eq. (\ref{eq:sum_sq_exp}), which is still valid if we replace $R_i$ by $Y_i$ and $x_i$ by $y_i$, the integral reads as follows
	\begin{equation}
		\label{eq_norm_integral}
		\begin{split}
			1 = & \, \mathcal{N}_{N}^{2} \ \kappa _{2}^{-1/2} \kappa _{1}^{1-N^2}\int e^{-Y_1^{2}} e^{-Y_{2n}^{2}} e^{-\sum\limits_{i=2}^{2N-1}Y_i^2}\\
			&\times \widetilde{V}(Y_1, \dots, Y_N)^2 \widetilde{V}(Y_{N+1}, \dots, Y_{2N})^2
			\ dY_{1} \cdots dY_{2N} \\
			1 = & \, \mathcal{N}_{N}^{2} \ \kappa _{1}^{1-N^{2}} \kappa _{2}^{-1/2} \ \mathcal{C}_{N}^{-2},
		\end{split}
	\end{equation}
	where $ \mathcal{C}_{N}$ is an interaction-independent constant. Therefore, we have that 
	\begin{equation}
		\mathcal{N}_N = \kappa_2^{1/4}  \kappa_1^{(N^2-1)/2} \mathcal{C}_N . 
	\end{equation}
	
	The integrand in the first line of Eq. \eqref{eq_norm_integral} is the square modulus of the ground-state wave function 
	\begin{equation}
		\psi_0 =  \, \mathcal{N}_N e^{-\frac{1}{2}R_1^2} \, e^{-\frac{\kappa_2}{2} R_{2N}^2} \, e^{-\frac{\kappa_1}{2} \sum\limits_{i=1}^{2N-1} R_i^2}
		\, \widetilde{V}(R_2, \dots, R_N)
		\, \widetilde{V}(R_{N+1}, \dots, R_{2N})
		\label{eq:sol_in_normal_modes}
	\end{equation}
	but taking $\kappa_1=\kappa_2=1$, i.e. $\Lambda=0$. In the last equation $\tilde{V}$ denotes the Vandermonde determinant in the $\{x_i\}$ coordinates but expressed in the $\{R_i\}$ variables. Notice also that this last equation is just another way of writing Eq. \eqref{eq_solucion}. 
	All  this means that $\mathcal{C}_N$ is precisely the corresponding value of $\mathcal{N}_N$ in the non-interacting regime. 
	
	For null interaction, the ground state is given by
	\begin{equation}
		\psi_0^{\Lambda=0} = \Psi_{\rm nif}(x_1, \dots , x_N)\, \Psi_{\rm nif}(x_{N+1}, \dots , x_{2N}),
	\end{equation}
	where
	\begin{equation*}
		\Psi_{\rm nif}(x_1, \dots , x_N)= A_N e^{-\frac{1}{2}\sum_{i=1}^{N} x_i^2} \, \prod_{1 \leq i<j \leq N} (x_i-x_j)
	\end{equation*}
	is the wave function of the ground state of $N$ identical non-interacting fermions in the harmonic trap \cite{girardeau2010two}, with the normalization factor
	\begin{equation}
		A_N = 2^{N(N-1)/4}\left[ N!\, \pi^{N/2}\, \prod_{k=0}^{N-1} k!\right]^{-1/2}.
	\end{equation}
	In summary, $\mathcal{C}_N = A_N^2$ and consequently
	\begin{equation}
		\mathcal{N}_N = \kappa_2^{1/4}\,  \kappa_1^{(N^2-1)/2}\, 2^{N(N-1)/2}\left[ N! \, \pi^{N/2}\, \prod_{k=0}^{N-1} k!\right]^{-1}.
	\end{equation}
	
	\section{Interacting fermions of the same kind}
	\label{sec_app_solutions_N_fermions}
	
	Consider the adimensionalized Hamiltonian for $N$ particles of kind $a$
	\begin{equation}
		H_a = \frac{1}{2}\sum _{i=1}^{N}\left[ -\frac{\partial ^{2}}{\partial x_{i}^{2}} + x_{i}^{2}\right] -\frac{\Lambda }{2}\sum _{i< j}( x_{i} -x_{j})^{2}.
	\end{equation}
	We can write
	\begin{equation}
		\sum _{1 \leq i< j \leq N}( x_{i} -x_{j})^{2}  =  \frac{1}{2}\sum _{i,j=1}^{N}( x_{i} -x_{j})^{2}
		= N\sum _{i=1}^{N} x_{i}^2 - \sum _{i,j=1}^{N} x_{i}x_{j},
	\end{equation}
	then the Hamiltonian can be written as
	\begin{equation}
		H_a=\frac{1}{2}\sum _{i=1}^{N}\left[ -\frac{\partial ^{2}}{\partial x_{i}^{2}} +( 1-N\Lambda ) x_{i}^{2}\right] +\frac{\Lambda }{2}\sum _{i,j=1}^{N} x_{i} K^{a}_{ij} x_{j},
	\end{equation}
	with
	\begin{equation}
		K^{a}=\left(\begin{array}{ c c c }
			1 & \cdots  & 1\\
			\vdots  & \ddots  & \vdots \\
			1 & \cdots  & 1
		\end{array}\right).
	\end{equation}
	This matrix has only one non-zero eigenvalue, which is equal to $N$, with the corresponding eigenvector
	\begin{equation}
		v^{(1)}=\frac{1}{\sqrt{N}}\left(1,\,1,\,\dots, 1\right)^T.
	\end{equation}
	We can use the orthonormal eigenbasis with eigenvalue zero
	\begin{equation}
		v_j^{( i)} =\frac{1}{\sqrt{i( i-1)}}\begin{cases}
			-1 & ,\ j< i\\
			i-1 & ,j=i\\
			0 & ,j >i
		\end{cases} \ \ \ \ \ \ \ \ \ i=2,...,N.
	\end{equation}
	By considering the variables
	\begin{equation}
		R_i=\sum_{j=1}^{N}v^{(i)}_j x_j,
	\end{equation}
	we obtain the solutions
	\begin{equation}
		\varphi_{n_1}\left(R_1\right)\varphi_{n_2}\left(\sqrt{\kappa_1}R_2\right)\cdots\varphi_{n_N}\left(\sqrt{\kappa_1}R_N\right),
	\end{equation}
	with $\kappa_1=\sqrt{1-N\Lambda}$, as previously defined. We can check that
	\begin{equation}
		\sum_{i=2}^{N}R_i^2 = \left( \sum_{i=1}^{N}x_i^2\right) - R_1^2=\frac{1}{N}\sum_{1 \leq i <j \leq N} \left(x_i-x_j\right)^2.
	\end{equation}
	This result has been obtained in \cite{benavides2014entanglement}. We get for the ground state of $N$ indistinguishable fermions
	\begin{equation}
		\psi_a= C_A e^{-\frac{1}{2N}\left(\sum_{i=1}^{N}x_i\right)^2} \prod _{1\leq i< j\leq N} e^{-\frac{\sqrt{1-N\Lambda }}{2N}( x_{i} -x_{j})^{2}}( x_{i} -x_{j}),
	\end{equation}
	where the normalization factor is given by
	\begin{equation}
		C_{A} =\kappa _{1}^{\left( N^{2} -1\right) /4} 2^{N(N-1)/4}\left[ N!\, \pi^{N/2}\, \prod_{k=0}^{N-1} k!\right]^{-1/2}.
	\end{equation}
	
	\section*{References}
	\bibliographystyle{iopart-num}
	\bibliography{Referencias}
\end{document}